\documentclass[aip, reprint,superscriptaddress]{revtex4-1}

\usepackage{booktabs}
\usepackage{mathrsfs,amssymb,graphics,subfigure,threeparttable}
\usepackage{graphicx}
\usepackage{epstopdf}
\usepackage{amssymb}
\usepackage{enumerate}
\usepackage{amsmath}
\usepackage{fancyhdr}
\usepackage{bm}

\usepackage[squaren]{SIunits}

\usepackage{url}
\PassOptionsToPackage{hyphens,spaces,obeyspaces}{url}
\usepackage[hidelinks,breaklinks]{hyperref}

\begin{document}

\title{A multi-sheath model for highly nonlinear plasma wakefields  }

\author{T. N. Dalichaouch}
\thanks{tdalichaouch@gmail.com}
\affiliation{Department of Physics and Astronomy, University of California, Los Angeles, California 90095, USA}

\author{ X. L. Xu}

\affiliation{SLAC National Accelerator Laboratory, Menlo Park, California 94025, USA}

\author{A. Tableman}
\affiliation{Department of Physics and Astronomy, University of California, Los Angeles, California 90095, USA}
\author{F. Li}
\affiliation{Department of Physics and Astronomy, University of California, Los Angeles, California 90095, USA}
\author{F. S. Tsung}
\affiliation{Department of Physics and Astronomy, University of California, Los Angeles, California 90095, USA}
\author{W. B. Mori}
\affiliation{Department of Physics and Astronomy, University of California, Los Angeles, California 90095, USA}
\affiliation{Department of Engineering, University of California, Los Angeles, California 90095, USA}

\date{\today}

\begin{abstract}

An improved description for nonlinear plasma wakefields with phase velocities near the speed of light is presented and compared against fully kinetic particle-in-cell simulations. These wakefields are excited by intense particle beams or lasers pushing plasma electrons radially outward, creating an ion bubble surrounded by a sheath of electrons characterized by the source term $S \equiv -\frac{1}{en_p}(\rho-J_z/c)$ where $\rho$ and $J_z$ are the charge and axial current densities. Previously, the sheath source term was described phenomenologically with a positive-definite function, resulting in a positive definite wake potential. In reality, the wake potential is negative at the rear of the ion column which is important for self-injection and accurate beam loading models. To account for this, we introduce a multi-sheath model in which the source term, $S$, of the plasma wake can be negative in regions outside the ion bubble. Using this model, we obtain a new expression for the wake potential and a modified differential equation for the bubble radius. Numerical results obtained from these equations are validated against particle-in-cell simulations for unloaded and loaded wakes. The new model provides accurate predictions of the shape and duration of trailing bunch current profiles that flatten plasma wakefields. It is also used to design a trailing bunch for a desired longitudinally varying loaded wakefield. We present beam loading results for laser wakefields and discuss how the model can be improved for laser drivers in future work. Finally, we discuss differences between the predictions of the multi- and single-sheath models for beam loading.

  \end{abstract}


\maketitle


\section{Introduction}

Research in plasma-based acceleration (PBA) driven by an intense laser pulse \cite{tajimadawson} or a relativistic particle beam \cite{chendawson1985} has been motivated by the capability to accelerate beams with gradients in excess of  a GeV/cm  over cm  or  larger  length scales \cite{hoganplasma4Gev10cm, blumenthal42GeV85cm, leemansplasma4GeV9cm, leemans20061GeV3cm, wang20132GeV7cm, hafz2008GeVbunches,litos2014high,adli2018acceleration,steinke2016multistage,PhysRevLett.122.084801}.  These wakefields can be excited by the space force of a particle beam (plasma wakefield acceleration--PWFA) or the radiation pressure of a laser (laser wakefield acceleration--LWFA). 
Such PBA based compact  accelerator stages could be the building blocks of next generation x-ray  free-electron-lasers (XFELs) or linear  colliders.

In PBA, a critical process is beam loading where a witness or trailing beam of particles is located at an appropriate phase of the wake where it is accelerated and focused. As it is accelerated it absorbs energy from the wake and can distort, i.e., load, it.  Developing an accurate beam loading theory is important in order to accurately understand and control the energy spread and emittance of the witness beam. In some cases, this needs to be understood even as the beam phase slips inside the wakefield. The witness beam can be externally or self-injected. Self-injection has advantages as it leads to synchronized injection which can be difficult to achieve for external injection due to the short periods and wavelengths of the plasma wakefields; however, self-injection may not produce the charge required for a linear collider. Recently, there have been many self-injection schemes proposed to generate high quality electron beams with low energy spread $\sigma_{\gamma}$ and normalized emittance $\epsilon_n$.  The most promising ideas typically involve decreasing the phase velocity $\gamma_{\phi}$ of the plasma wake using either a plasma density down ramp \cite{katsouleasdownramp1986,bulanovdownramp1998,sukdownramp2001, xu2017downrampinj,martinezdownramp2017} or an evolving driver \cite{kalmykovevolvingplasmabubble2011, xu2005tightfocusedlaser, Thamine2020}. In each of these instances, plasma electrons are injected at the very rear of the first bucket of the wake where they can then be accelerated over long periods of time. 

In order to  characterize how the injected beams alter the wakefield, a theoretical model for the wakefield that is accurate in the rear of the  region is required. If the model is accurate enough it can also be used to design experiments and simulations capable of generating injected beams that can flatten a wakefield or provide the necessary slope in the acceleration gradient to compensate for an initial energy chirp after some acceleration distance.

\begin{figure}[t]
\includegraphics[width=0.5\textwidth]{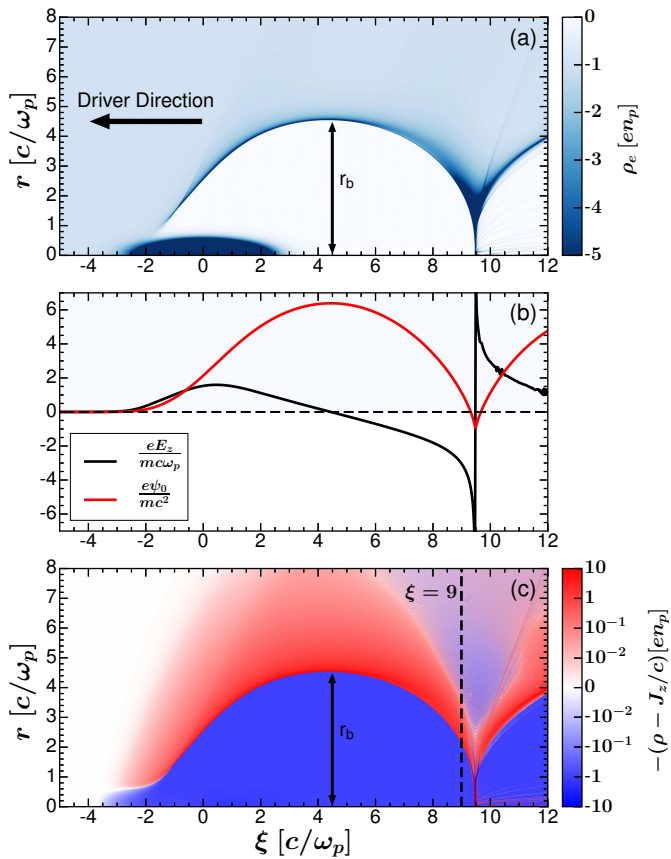}
\caption{\label{fig:schematic1} (a) Electron density distribution of a plasma wake excited by an electron drive beam with parameters $\Lambda_d = 6$, $\gamma_b = 20000$, $k_p\sigma_z = 1$, $k_p\sigma_r =0.245$, and $k_p \xi_c = 0$. The maximum bubble radius is $k_pr_m \simeq 4.53$. (b) The on-axis electric field $E_z$ and wake potential $\psi_0$. (c) Contour plot of the source term $-(\rho - J_z/c)$.  }
\end{figure}

In the linear regime, the necessary beam loading theory has existed for over thirty years \cite{Katsouleas87}. However, in the nonlinear regime the theory is significantly more complicated. In nonlinear wakefields the plasma electrons are expelled by the space-charge force of a particle beam (PWFA) or radiation pressure of a laser pulse (LWFA) leaving behind a column of ions. These electrons, which are initially blown-out, are attracted back to the axis due to the space-charge force from the ions, forming a plasma sheath covering a nearly spherical ion channel radius $r_b(\xi)$. This structure can be seen in Fig.~\ref{fig:schematic1}(a), where the electron density from a PWFA simulation using the particle-in-cell (PIC) code {\scshape osiris} \cite{osiris} is plotted. A non-evolving driver with a peak normalized charge per unit length $\Lambda \equiv 4\pi r_e \int^{r\gg \sigma_r}_0  n_b r dr=6$, energy $\gamma_b =20000$, spot size $k_p\sigma_r = 0.245$, duration $k_p \sigma_z =1$, and centroid $k_p\xi_c = 0$ was used, where $k_p = \omega_p/c$ is the plasma wavenumber, $\omega_p^2 = \frac{4 \pi e^2 n_p}{m}$ is the plasma frequency, and $n_b$ is the drive beam density. In seminal papers by Lu et al.~\cite{lu2006nonlineartheoryprl, lu2006nonlinearphysplasma}, a nonlinear theory was introduced to characterize the structure and fields generated by these kinds of three-dimensional plasma wakes operating in the blowout regime. Using a co-moving coordinate $\xi \equiv (ct-z)$ and the quasi-static approximation, it was shown that expressions for the electric and magnetic fields of the  wake inside the ion column, as well as a differential equation for the  bubble trajectory $r_b(\xi)$, could be determined for given models for the sheath.  Tzoufras et al.~\cite{tzoufrasprl, tzoufrasprab} showed that in the nonlinear regime  beam loading  arises through modifications to $r_b(\xi)$ from the electromagnetic forces of the witness beam.

All of the forces on a particle moving close to the speed of light can be obtained from the normalized wake potential $\psi = \frac{e}{mc^2}(\phi - A_z)$, where $\phi$ is the electric potential and $A_z$ is the  vector potential in the $\hat z$ direction. In Refs.~\citenum{lu2006nonlineartheoryprl} and \citenum{lu2006nonlinearphysplasma}, it was shown the wake potential can be obtained from a two dimensional Poisson equation 

\begin{align}
\label{eq:delpsi}
\nabla_{\perp}^2 \psi =S \equiv -\frac{1}{en_p}(\rho - J_z/c)
\end{align}
where the integral of the source term over the transverse coordinates in each $\xi$ slice vanishes. 

For azimuthally symmetric beams or laser drivers, the solutions to this equation for radii inside the ion column have the form $\psi=\frac{1}{4}(1+\beta)r_b^2-r^2$ where $\beta$ depends on integrals over transverse gradients of $\psi$ written in terms of $S$ from $r=\infty$ to  0. It was shown in Refs.~\citenum{lu2006nonlineartheoryprl} and \citenum{lu2006nonlinearphysplasma} that for a source term comprised of two regions where $S = -1$ inside the ion bubble $(r < r_b)$ and $S \geq 0$ in a finite width plasma sheath outside the bubble $( r_b <r < r_b+ \Delta)$, $\beta = \frac{(1+\alpha)^2 \ln(1+\alpha)^2}{(1+\alpha)^2-1} -1$, where $\alpha = \frac{\Delta}{r_b}$. Using this expression for $\beta$  a differential equation for $r_b$ was then obtained (see equation 46 in Ref.~\citenum{lu2006nonlinearphysplasma}). It was shown that these equations could explain many of the features observed in particle-in-cell simulations.
 
However, this simple model for the plasma sheath and hence the wake potential has its limitations. For example, a 
direct consequence of using such a model is that $\beta$  is positive definite for each $\xi$ slice and hence $\psi$ is positive definite at all locations within the ion column.

In one dimension, wavelike analysis to the cold fluid equations show that solutions exist until wavebreaking occurs. This can be  physically interpreted as the limit where the plasma density compression approaches infinity, the electric field fully steepens (its slope approaches infinity), two plasma sheets cross,  particle trapping of a background electron occurs, i.e, a particle moves with the phase velocity of the wave \cite{mori90}. In this case,  the minimum wake potential becomes $-1$ which is the threshold for particle trapping. Even in multi-dimensions, the trapping condition for background electrons, $\psi < -1+\frac{(1+p_\perp/mc)^2}{\gamma_\phi}$  \cite{ionizationinjconcept2006-3}, also requires  that the wake potential approach -1. PIC simulations show empirically that the wake potential is negative in the rear of the first bucket as shown in Fig.~\ref{fig:schematic1}(b). In this region of the wake, plasma sheath electrons can be accelerated to large forward velocities $\gamma_z \equiv (1-\beta_z^2)^{-1/2} \gg 1$ as they return back to the $\xi$-axis due to the large accelerating fields $E_z(\xi)$ at the rear of the wake. Near the axis, the wake potential $\psi$ must approach $-1$ if $v_z$ approaches c as can be seen from the constant of motion equation $\gamma - P_z/mc = 1+\psi$ \cite{mora}. In fact, many self-injection schemes \cite{ katsouleasdownramp1986, bulanovdownramp1998,sukdownramp2001, xu2017downrampinj,martinezdownramp2017,kalmykovevolvingplasmabubble2011, xu2005tightfocusedlaser, Thamine2020} rely on $\psi$ approaching $-1$ at the rear of the wake in order to satisfy the electron trapping condition $\gamma_{z} > \gamma_{\phi} \gg 1$. Therefore, while the model used by Lu et al.~\cite{lu2006nonlinearphysplasma,lu2006nonlineartheoryprl} can predict the bubble trajectory $r_b(\xi)$ and longitudinal electric field $E_z(\xi)$ in regions where the wake potential $\psi(\xi)$ is sufficiently positive, it will not be accurate near the rear of the bubble where the wake potential becomes negative.

As noted in Ref.~\citenum{xu2017downrampinj}, if the source term $S$ is negative in some region outside the bubble $r > r_b$ then wake potential $\psi$ can be  negative inside the ion column. This is illustrated in Fig.~\ref{fig:schematic1}(c) where three distinct regions are evident from the contour plot of $S$. In addition to the ion channel and plasma sheath regions included in the simple single-sheath model employed in Refs.~\citenum{lu2006nonlinearphysplasma} and \citenum{lu2006nonlineartheoryprl}, it is clear that there is a third region of finite width where $S < 0$ at the rear of the wake outside the bubble $r > r_b$. This negative region is highly localized to the rear of the bubble and drops off rapidly in terms of amplitude at $\xi$ where the bubble radii is larger.

In this manuscript, we propose to use a multi-sheath model for the source term $S$ comprised of three regions to obtain a new expression for the wake potential $\psi(\xi)$. Using the proposed model for $\psi(\xi)$ in conjunction with the nonlinear blowout theory presented by Lu et al.~\cite{lu2006nonlineartheoryprl, lu2006nonlinearphysplasma}, we will calculate the trajectory of the bubble radius $r_b(\xi)$ and electric field $E_z(\xi)$ while using the constants of motion to constrain the variables. We find that numerical results obtained using the proposed model agree well with PIC simulation results throughout the entire ion column. We also  compare the results for the multi-sheath model to the those from the single  sheath model employed in Refs.~\citenum{lu2006nonlinearphysplasma} and \citenum{lu2006nonlineartheoryprl}.  We also show the importance of using the multi-sheath model when studying beam loading of nonlinear wakes from witness electron beams. To accurately analyze beam loading in the nonlinear regime it is essential to have an accurate equation for  $r_b(\xi)$ and  for $\psi(r<r_b)$. In the original work of Tzoufras et al.~\cite{tzoufrasprl, tzoufrasprab}, beam loading was analyzed by determining how $r_b(\xi)$ is modified by the electromagnetic forces  of the witness beam. To obtain analytical results, Tzoufras et al.~applied the large $r_b$ limit to  the differential equation for $r_b$. We show that the multi-sheath model provides better agreement and apply it to linear collider and self-injection parameters. Current profiles of witness beams that flatten the wakes are provided. We will show numerical results obtained for the multi-sheath model for wakes excited by intense lasers. The theory does not work as well as the laser driven wakefields do not lead to complete blowout and more complicated sheath structures. Last, a detailed discussion on the differences between the results for the multi-sheath and single sheath models is given.

\section{The plasma wake potential}
\label{sec:potential}
The goal of this work is to obtain a more accurate expression for the differential equation for $r_b(\xi)$ and the fields inside the ion column. We begin by concentrating on the wake potential from which the accelerating and focusing fields for a witness beam are derived.  The differential equation for $r_b(\xi)$ can therefore be completely described if the wake potential is known. As mentioned above, the simple sheath model used in Refs.~\citenum{lu2006nonlineartheoryprl} and \citenum{lu2006nonlinearphysplasma} cannot accurately describe $r_b$ or the wake potential unless $k_p r_b (\xi)$ is sufficiently large. For such a simple sheath model, the wake potential is positive definite; however, it is known empirically from PIC simulations that the wake potential approaches $-1$ at the rear of nonlinear multi-dimensional wakes. In subsequent sections, we show that in order to get accurate predictions for $r_b$ and beam loading it is essential that the potential approach $-1$. 
 
\begin{figure}[t]
\includegraphics[width=0.5\textwidth]{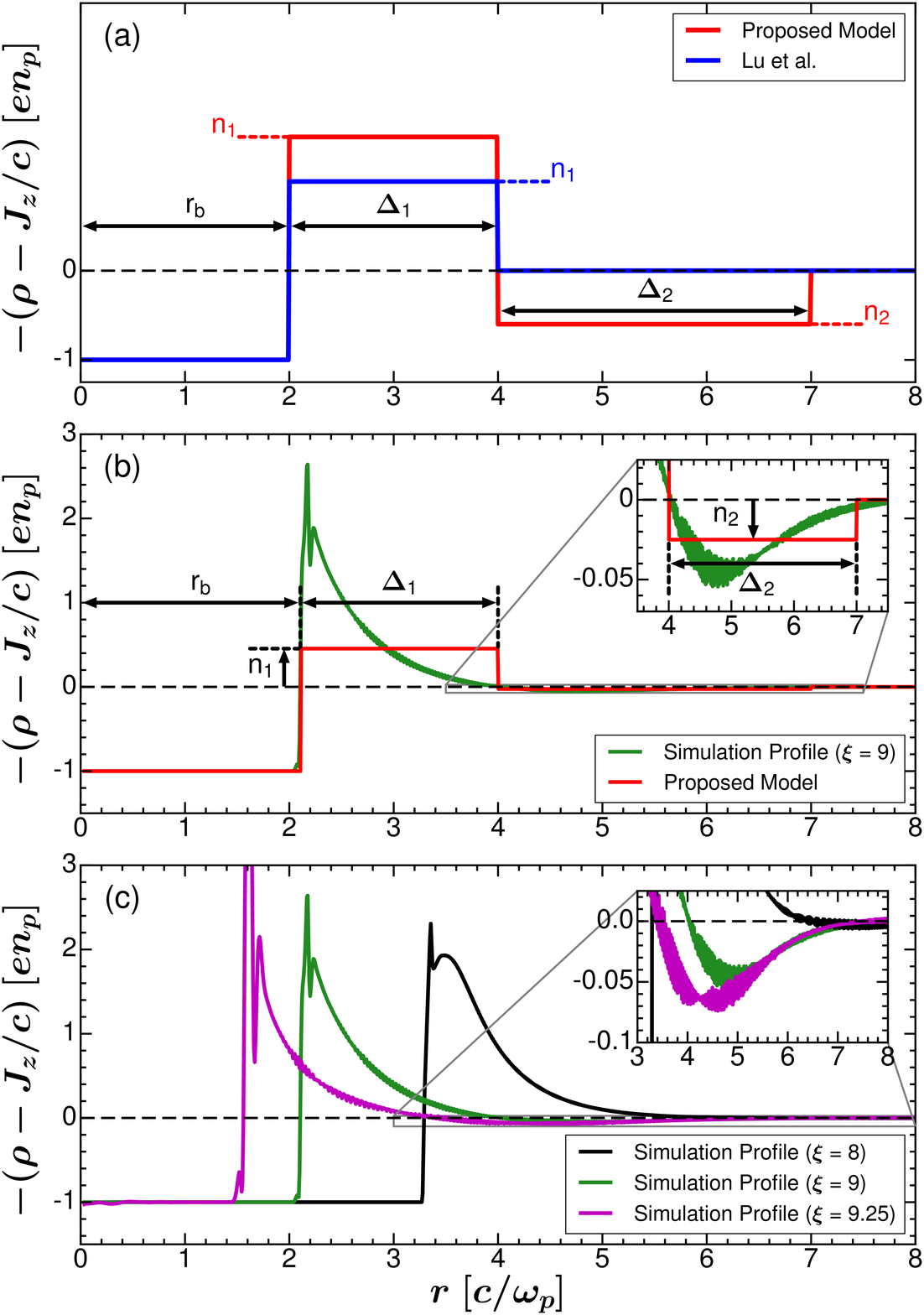}
\caption{\label{fig:schematic2} (a) Comparison of the proposed multi-sheath model profile (red) and single-sheath model \cite{lu2006nonlinearphysplasma,lu2006nonlineartheoryprl} (blue). (b)  The proposed model profile (red) and simulation profile (green) for the transverse slice at $k_p\xi = 9$ (dashed black) from Fig.~\ref{fig:schematic1}(c). (c) Transverse slices of the source term profiles from Fig.~\ref{fig:schematic1}(c) at different $k_p\xi $.  }
\end{figure}

As noted above, it is straightforward to show that in order for $\psi$ to be negative the source term must be negative for some $r>r_b$ beyond the sheath. We thus propose a phenomenological source term $S$ that extends the single-sheath model from Refs.~\citenum{lu2006nonlinearphysplasma} and \citenum{lu2006nonlineartheoryprl} by introducing a second plasma sheath $\Delta_2$ in which the source term is negative, i.e., $S \equiv n_2 < 0$, outside the ion bubble $r> r_b$. This is shown schematically in Fig.~\ref{fig:schematic2}(a). For comparison, we also show the simple model utilized by Lu et al. \cite{lu2006nonlineartheoryprl, lu2006nonlinearphysplasma}. While two sheaths are usually enough to model the source term for wakes created by electron beams (PWFA problems), the formalism can be  extended to include an arbitrary number of sheaths. This may be needed for accurate descriptions of nonlinear wakes created by laser drivers which will be discussed later. Therefore, we refer to the proposed model as the ``multi-sheath model" since it can employ two or more sheaths while we refer to model used by Lu et al.~\cite{lu2006nonlinearphysplasma,lu2006nonlineartheoryprl} as the ``single-sheath model" since it employs only one sheath. 

As we will show in this section, this second sheath region $\Delta_2$ is needed to describe both the physics and mathematics of the plasma wake features at the rear of the bubble. Once an expression for $\psi$ is obtained using the proposed model for $S$, we can solve for the trajectory of the bubble radius $r_b$, $\psi$, and accelerating field $E_z$.  We note that others \cite{Shvets, Mehrling}
have proposed different phenomenological sheath models than the one presented here. These authors were motivated to obtain accurate descriptions for the fields inside the sheath in order to study self-injection and hosing.  However, these sheath models do not address the shortcomings described here and, in some cases, they are also not necessarily self-consistent in that they do not conserve charge within each slice. They thus cannot properly address self-injection.

Henceforth,  we will employ normalized units, where charge is normalized to electron charge $e$, mass to electron mass $m$, velocity to $c$, charge density to $en_p$, current density to $en_pc$, length to $c/\omega_p$, time to $\omega_p^{-1}$, electric fields to $mc\omega_p/e$, and potentials to $mc^2/e$. We will also assume that the wake is excited by a bi-Gaussian electron bunch with a density profile $n_b \sim e^{-r^2/(2\sigma_r^2)}e^{-\xi^2/(2\sigma_z^2)}$ and a spot size $\sigma_r$ much smaller than the blowout radius $r_m$. Following the convention used in Refs.~\citenum{lu2006nonlinearphysplasma} and \citenum{lu2006nonlineartheoryprl}, we will use step functions to model the source term $S$ as illustrated in Fig.~\ref{fig:schematic2}(a) with
\begin{equation}
\label{eq:sprofile}
    S= 
    \begin{cases}
     -1, & \text{if}\ r < r_b \\
      n_1, & \text{if}\ r_b <r < r_b+ \Delta_1 \\
      n_2, & \text{if}\ r_b+\Delta_1< r < r_b+ \Delta_1+\Delta_2 \\
      0, & \text{otherwise.}
    \end{cases}
\end{equation}

In Fig.~\ref{fig:schematic2}(b), we show how the multi-sheath model compares to the actual source term profile at $k_p\xi = 9$ obtained from the PIC simulation in Fig.~\ref{fig:schematic1}. In order to understand the physics represented in each region we write out the electron and ion source terms, $S = -(\rho - J_z) = -\rho_{ion} - \rho_e(1 - v_z)$. Inside the bubble,   $r<r_b$ (region \text{I}), $\rho_e=0$ and $\rho_{ion}=1$ so $S (r<r_b)=-1$ as shown in Fig.~\ref{fig:schematic1}(a). Due to space-charge separation from blowout, plasma electrons are attracted back to the $\xi$-axis by the ion channel, thereby forming a plasma sheath (region \text{II}) with a large negative density spike $\rho_e \ll -1$ at the bubble interface $r_b(\xi)$. From the constant of motion for a plasma particle $\bar{\gamma} - P_z=1+\psi $ \cite{mora}, it can be shown that $v_z = \frac{1+P_{\perp}^2 - (1 +\psi)^2}{1+P_{\perp}^2 + (1+\psi)^2}$ \cite{lu2006nonlinearphysplasma,lu2006nonlineartheoryprl}. Therefore, the innermost sheath electrons at the top of bubble, where $P_{\perp} \simeq 0$, propagate backwards $v_z < 0$. Since $(1-v_z) > 1$ and $\rho_e \ll -1$ in this region, the innermost sheath is characterized by a positive source term $S \equiv n_1 > 0$ within a finite width region denoted by $\Delta_1$. As these innermost sheath electrons return back to the axis where the wake potential $\psi$ can approach $-1$, they can propagate in the forward direction at nearly the speed of light, i.e., $v_z \sim 1$. In this region, the source term of the electrons is reduced by the factor $(1-v_z) \ll 1$. Despite this, the sheath width $\Delta_{1}$ remains finite at the rear of the wake because the electron density spike along the bubble interface $r_b(\xi)$ is large enough to offset the ion term, i.e., $-\rho_e  > \frac{1}{(1-v_z)} \gg 1$.

Near the back of the bubble, there exists a second plasma sheath (region \text{III}) of width $\Delta_2$ bordering the first in which the source term $S \equiv n_2 < 0$. In this region, the electron density is of the order of unity, i.e., $\rho_e \lesssim -1$, and plasma electrons are still propagating forward, i.e., $v_z > 0$. Therefore, $-\rho_e(1-v_z) < 1$, resulting in a negative source term $S < 0$. 

As can be seen in Fig.~\ref{fig:schematic2}(c), the simulation profile of the second sheath varies with the bubble radius $r_b(\xi)$ and $\xi$. For $r_b(\xi)$ close to the maximum radius $r_m \simeq 4.53$, the second sheath can generally be neglected since $n_2 \approx 0$. However, the amplitude of $n_2$ rapidly increases as $r_b(\xi)$ decreases, which is observed for transverse slices located at $k_p\xi = 9, 9.25$ in Fig.~\ref{fig:schematic1}(c). In most simulations, peak $n_2$ values of $\sim O(-\frac{1}{10})$ can be observed in the second sheath region at the very rear of the wake where $r_b (\xi)$ goes to zero. Despite the fact that $n_2$ is small, the width $\Delta_2$ usually extends over several plasma skin depths. Therefore, the source term of the second sheath actually contributes the most to the negative pseudopotentials observed at the rear of the wake when integrating Eq.~(\ref{eq:delpsi}). In regions $r \gg r_b$ far from the blowout, both $\Delta_1$ and $\Delta_2$ connect to the linear regime where electron perturbation $|\delta\rho_e| \ll 1$ is small and the electron velocities $|v_z| \ll 1$ are non-relativistic. In this limit, electrons oscillate at the plasma frequency $\omega_p$.

The parameters defined in Eq.~(\ref{eq:sprofile}), $n_1, n_2, \Delta_1$, and  $\Delta_2$, are related by the requirement that charge is conserved in each slice as derived in Refs.~\citenum{lu2006nonlinearphysplasma} and \citenum{lu2006nonlineartheoryprl},

\begin{align}
\label{eq:constant1}
\int^{\infty}_0 (\rho - J_z) rdr = 0.
\end{align}
Integrating Eq.~(\ref{eq:constant1}), we obtain  

\begin{align}
\label{eq:constant1_eval}
\begin{split}
&-1 + n_1\left[ \left(1+\frac{\Delta_1}{r_b}\right)^2 -1 \right] +  \\ 
& n_2 \left[ \left(1+\frac{\Delta_1+\Delta_2}{r_b}\right)^2 - \left(1+ \frac{\Delta_1}{r_b}\right)^2 \right] = 0.
\end{split}
\end{align}
We can rewrite Eq.~(\ref{eq:constant1_eval}) to solve for $n_1$ in terms of $n_2$, $\alpha_1$, and $\alpha_2$, 
\begin{align}
\label{eq:constant1_eval2}
\begin{split}
n_1 = \frac{1-n_2\left(\alpha_2^2 + 2\alpha_2\alpha_1 + 2\alpha_2 \right)}{ \left(1+\alpha_1\right)^2 -1} 
\end{split}
\end{align}
where $\alpha_{1} \equiv \frac{\Delta_{1}}{r_b}$ and $\alpha_{2} \equiv \frac{\Delta_{2}}{r_b}$. If $n_2=0$, we can recover the expression $n_1 = n_{\Delta} \equiv \frac{1}{(1+\alpha_1)^2 - 1}$ from the single-sheath model.

We next calculate the wake potential $\psi(r, \xi)$. To do this, we first need to determine the on-axis potential $\psi_0(\xi) \equiv \psi(0,\xi)$. Once $\psi_0(\xi)$ is known, the wake potential $\psi(r,\xi) = \psi_0(\xi) - \frac{r^2}{4}$ is defined everywhere inside the bubble . To obtain $\psi_0(\xi)$, we integrate Eq.~(\ref{eq:delpsi}) across all three regions defined in Eq.~(\ref{eq:sprofile})
\begin{align}
\label{eq:psi_integral}
\psi_0 &= \int^{\infty}_0 \frac{dr}{r} \int^r_0 (\rho - J_z) r^{\prime} dr^{\prime} \notag \\
&=\left[ \int^{r_b}_0 +  \int^{r_b+\Delta_1}_{r_b} +  \int^{r_b + \Delta_1 + \Delta_2}_{r_b + \Delta_1} \right]\frac{dr}{r} \int^r_0 (-S) r^{\prime} dr^{\prime}  \notag \\
&=\Psi_{\text{I}} +  \Psi_{\text{II}} + \Psi_{\text{III}}
\end{align}
where $\Psi_\text{I} = \frac{r_b^2}{4}$ is the contribution from the ion bubble, $\Psi_\text{II} = \frac{r_b^2}{4}\bigg\{ 2(1+n_1) \ln(1+\alpha_1) -n_1\bigg[ (1+\alpha_1)^2 - 1\bigg] \bigg\}$ is the contribution innermost plasma sheath of width $\Delta_1$, and $\Psi_\text{III} = \frac{r_b^2}{4} \bigg\{ 2 n_2(1+\alpha_1+\alpha_2)^2 \ln\left(1 + \frac{\alpha_2}{1+\alpha_1}\right) -n_2 \bigg[ (1+\alpha_1 +\alpha_2)^2- (1+\alpha_1)^2 \bigg] \bigg\}$ is the contribution from second plasma sheath of width $\Delta_2$. 

Summing the expressions of all the three regions and simplifying with Eq.~(\ref{eq:constant1_eval2}), we obtain a final expression for the wake potential inside the bubble $(r \leq r_b(\xi))$, similar to the one derived in Refs.~\citenum{lu2006nonlineartheoryprl} and \citenum{lu2006nonlinearphysplasma},

\begin{align}
\label{eq:psi_final}
\psi(r,\xi) &= \psi_0(\xi) - \frac{r^2}{4} \notag \\
&= \frac{r_b^2 (\xi)}{4} (1+\beta^{\prime}) - \frac{r^2}{4}
\end{align}
where 
\begin{align}
\label{eq:beta}
\beta^{\prime} &= 2(1+n_1)\ln(1+\alpha_1) -1 \notag \\
&+ 2n_2(1+\alpha_1 + \alpha_2)^2 \ln\left(1+\frac{\alpha_2}{1+\alpha_1}\right).
\end{align}

Eqs.~(\ref{eq:psi_final}) and (\ref{eq:beta}) contain the key differences between the present work and that in Refs.~\citenum{lu2006nonlineartheoryprl} and \citenum{lu2006nonlinearphysplasma}. Naturally, these differences also effect the bubble trajectory $r_b$ since the plasma forces depend on $\psi$. Thus, it is worth comparing and discussing the differences between $\beta$ and $\beta'$. First, if $n_2$ is set to zero then it is trivial to see that $\beta'$ reduces to $\beta \equiv \frac{(1+\alpha_1)^2 \ln(1+\alpha_1)^2}{(1+\alpha_1)^2-1} -1$ (using the conservation of charge to relate get $n_1$ as a function of $\alpha_1$) which is only a function of $\alpha_1$, and we recover the on-axis wake potential $\psi_0= (1+\beta)r_b^2/4$. Furthermore, it is important to note that  $r_b^2\beta\rightarrow 0$ as $r_b\rightarrow 0$ and therefore the minimum $\psi$ at rear of the bubble is 0 for the single-sheath model. 

On the other hand $\beta^{\prime}$ is a function of four parameters $\beta^{\prime}(n_1, n_2, \alpha_1, \alpha_2)$ where each of these parameters are unknown functions of $r_b$. The goal is to use a combinations of physics constraints and phenomenological arguments to reduce $\beta^{\prime}$ to be a known function of $r_b$. First, we use the conservation of charge constraint, Eq. (\ref{eq:constant1_eval2}), which gives $n_1(n_2, \alpha_1, \alpha_2)$ to eliminate $n_1$ from $\beta^{\prime}$. Next, we will use empirical observations regarding $\psi_0 (r_b=0)$ and phenomenological arguments for the dependence of  $n_2$, $\Delta_1$, and $\Delta_2$ to obtain an expression for $\beta^{\prime}$ in terms of $r_b$.

We assume that the sheath widths are finite as $r_b$ approaches 0 and can thus we written as $\Delta_1(r_b)=\Delta_{10} c_1(r_b)$,  $\Delta_2(r_b)=\Delta_{20} c_2(r_b)$, and $n_2=n_{20}h(r_b)$ where $c_1$,  $c_2$, and $h$ are functions of $r_b$ that approach 1 as $r_b\rightarrow 0$. We next  take the limit of $\psi(r_b)$ as $r_b \rightarrow 0$ to obtain a relationship between the empirical value of $\psi_{min} \equiv \lim_{r_b\rightarrow 0} \psi(r_b(\xi))$, and $\Delta_{10}$, $\Delta_{20}$, and $n_{20}$,

\begin{align}
\label{eq:psi_limit}
\psi_{min} & \equiv \lim_{r_b(\xi)\rightarrow 0}\beta^{\prime}\frac{r_b^2}{4} \notag \\
&=   \lim_{r_b(\xi)\rightarrow 0}   \bigg[ (1+  n_1) \frac{r_b^2}{2} \ln \left(1 + \frac{\Delta_1}{r_b} \right)-\frac{r_b^2}{4}  \notag \\
&+ n_2( r_b +  \Delta_1 + \Delta_2)^2 \ln\left(1 + \frac{\Delta_2}{r_b + \Delta_1} \right) \bigg] \notag \\
&=  \frac{n_{20}}{2}   (  \Delta_{10} + \Delta_{20})^2 \ln\left(1+\frac{\Delta_{20}}{\Delta_{10}}\right).
\end{align}
It is straightforward to show that the first term in the limit vanishes since $\lim_{r_b\rightarrow 0} r_b^2 \ln(1+ \Delta_1/r_b) = 0$ and the amplitude of the innermost plasma sheath remains finite $n_{10} \equiv n_1(r_b = 0) = \frac{-n_{20} (\Delta_{20}^2  + 2 \Delta_{20}\Delta_{10} + 2\Delta_{20})}{\Delta_{10}^2}$ due to charge conservation with the second sheath from Eq.~(\ref{eq:constant1_eval2}). Rearranging the terms in Eq.~(\ref{eq:psi_limit}) to solve for $n_{20}$, we obtain

\begin{align}
\label{eq:n2_limit}
n_{20} = \frac{2\psi_{min}}{ (\Delta_{10} + \Delta_{20})^2 \ln\left(1+\frac{\Delta_{20}}{\Delta_{10}}\right)}
\end{align}
where $\psi_{min}$ is the minimum value of the on-axis potential $\psi_0$. Empirically, it is known that $\psi_{min}$ is negative, from which it follows from Eq.~(\ref{eq:n2_limit}) that $n_{20}$ must also be negative, which had been argued above when we motivated the need for the multi-sheath model. Additionally, from Eq.~(\ref{eq:constant1_eval2}), it also follows that $n_{10}$ must be positive if $n_{20}$ is negative.

Eq.~(\ref{eq:n2_limit}) is important because it constrains the parameter $n_{20}$ for given value of $\Delta_{10}, \Delta_{20}$, and $\psi_{min}$. While PIC simulations can be used to determine the exact value of $\psi_{min}$ in the nonlinear blowout regime, $\psi_{min}$ can be well approximated by $-1$ when the maximum bubble radius is sufficiently large, i.e., $r_m \gtrsim 3 $. Under these conditions, sheath electrons that trace the bubble $r_b(\xi)$ travel near the speed of light, $v_z \sim 1$, with finite transverse momentum $P_{\perp}$ at the rear of the wake \cite{xu2017downrampinj,Thamine2020}. From the constant of motion for a plasma particle $\bar{\gamma} - P_z=1+\psi $ \cite{mora}, it can be shown that $1-v_z = \frac{2(1+\psi)^2}{1+P_{\perp}^2 + (1+\psi)^2}$ \cite{lu2006nonlinearphysplasma,lu2006nonlineartheoryprl} so that $v_z\rightarrow 1$ when $\psi_{min} \rightarrow -1$.

Until this point, we have not specified $n_1(r_b)$, $n_2(r_b)$, $\Delta_1(r_b)$ and $\Delta_2(r_b)$. In general, since the phenomenological model for $S$ employs simple step functions in each region, we will not be able to fit these parameters exactly to empirical wake structures across all regions. Instead, the goal is to use profiles for $n_1$, $n_2$, $\Delta_1$ and $\Delta_2$ that can reproduce the on-axis pseudopotential $\psi_0(\xi)$ and innermost electron trajectory $r_b(\xi)$ for nonlinear plasma wakes. Once these quantities are determined, the wake potential $\psi(r,\xi) = \psi_0(\xi) - \frac{r^2}{4}$ would be correct everywhere inside the bubble $r \leq r_b(\xi)$, which is the region of interest in the nonlinear blowout regime.

Following the single-sheath model~\cite{lu2006nonlinearphysplasma,lu2006nonlineartheoryprl}, we use a profile for the first plasma sheath width of $\Delta_1 = \Delta_{10} + \Delta_s$, where $\Delta_{10} \sim O(1)$ and $\Delta_s =\epsilon r_b$ such that $c_1=1+\epsilon r_b$. This profile is consistent with the physical picture described earlier in which $\Delta_{1}$ remains finite near the axis due to the large electron density spike $|\rho_e| \gg 1$ along $r_b(\xi)$. It is worth pointing out that the values of $\Delta_{10}$ and $\epsilon$ can be varied slightly to adjust the length of the plasma wake obtained by integrating Eq.~(\ref{eq:rbeq}). We use $\Delta_{10}=1$ and $\epsilon=.05$ in most cases and explicitly assume they do not depend on $\xi$.

To model the $S < 0$ region outside the bubble, we use a Gaussian profile $n_2 = n_{20} e^{-s r_b^2/r_m^2}$ such that $h=e^{-s r_b^2/r_m^2}$ and a constant width $\Delta_{2} = \Delta_{20}$ for simplicity, where $r_m$ is the maximum value of $r_b$. While other profiles can be used to model this region, the Gaussian profile is largely motivated by the observed behavior of $n_2$ in PIC simulations [see Fig.~\ref{fig:schematic2}(c)] where it reaches a negative minimum when $r_b = 0$ and approaches 0 as $r_b$ approaches the blowout radius $r_m$.  We note that super Gaussians can be used, $h=e^{-s r_b^t/r_m^t}$ where tuning $t$ can improve the accuracy. The value of $n_{20}$ is obtained using Eq.~(\ref{eq:n2_limit}), where the minimum wake potential $\psi_{min}$ also needs to be provided as input. For nonlinear wakes in the blowout regime, i.e., $r_m \gtrsim 3$, $\psi_{min} \approx -1$ can be used. For all cases presented in this paper, we use $s=3$ for the Gaussian coefficient $(t=2)$ and $\Delta_{20} =3$ for the second sheath so that the $n_{20}$ values calculated from Eq.~(\ref{eq:n2_limit}) are in agreement with values observed in simulations that are typically $\sim O(-\frac{1}{10})$. However, the results are largely insensitive to values of $s$ and $\Delta_{20}$ ranging from 2 to 4.

We can now use $\beta^{\prime}(n_{20}, \Delta_{10}, \Delta_{20}, h(s, r_b), c_1(\epsilon, r_b)/r_b)$ (note that $c_2=1$) to obtain the trajectory of the bubble $r_b$ for a given $r_m$. In a plasma wake, the bubble radius $r_b$ starts at 0 reaches a maximum value of $r_m$ and then returns to 0 at the rear. In the single-sheath model, $\beta(r_b)$ is single valued (symmetric) inside the bubble and the sheath structure looks the same in the front and back half of the bubble. On the other hand, in the multi-sheath model (and in real wakes), the sheath structure looks different between the front and back half (there is a single sheath in the front half). While it may be possible to merge the two models, in what follows we concentrate on examining the back half of the bubble for both loaded and unloaded wakes starting from the maximum blowout radius $r_b = r_m$. 

As noted above much of the formalism in \cite{lu2006nonlinearphysplasma,lu2006nonlineartheoryprl}  is independent of the choice for the sheath model.  Following the same procedure, the differential equation describing the trajectory of the innermost plasma electron tracing the ion channel $r_b(\xi)$ can be shown to be,

\begin{align}
\label{eq:generalrb}
\frac{d}{d\xi} \Bigg[ (1+\psi) \frac{d}{d\xi}r_b \Bigg] &= r_b \Bigg\{ -\frac{1}{4} \Bigg[1 + \frac{1}{(1+\psi)^2} + \left(\frac{dr_b}{d\xi}\right)^2 \Bigg] \notag \\
&- \frac{1}{2} \frac{d^2\psi_0}{d\xi^2}  + \frac{\lambda(\xi)}{r_b^2}   \Bigg\}
\end{align}
where the current profile of the drive and/or trailing bunch is given by $\lambda(\xi) = \int^{r\gg \sigma_r}_0 n_b rdr$. By expressing $\psi(r,\xi)$ in the form shown in Eq.~(\ref{eq:psi_final}), we can obtain a differential equation for the innermost particle trajectory $r_b(\xi)$ 
\begin{align}
\label{eq:rbeq}
A^{\prime}(r_b) \frac{d^2r_b}{d\xi^2} + B^{\prime}(r_b) r_b \left( \frac{dr_b}{d\xi}\right)^2 + C^{\prime}(r_b)r_b = \frac{\lambda (\xi)}{r_b}
\end{align}
where the coefficients $A^{\prime}(r_b)$, $B^{\prime}(r_b)$, and $C^{\prime}(r_b)$ are

\begin{align*}
A^{\prime}(r_b)  &= 1+ \left[\frac{1}{4} + \frac{\beta^{\prime}}{2} + \frac{1}{8} r_b \frac{d\beta^{\prime}}{dr_b} \right]r_b^2, \notag \\
\notag \\
B^{\prime}(r_b)  &= \frac{1}{2} + \frac{3}{4}\beta^{\prime} + \frac{3}{4} r_b \frac{d\beta^{\prime}}{dr_b}  + \frac{1}{8} r_b^2 \frac{d^2\beta^{\prime}}{dr_b^2}, \notag \\
\notag \\
C^{\prime}(r_b)  &= \frac{1}{4} \left[\vcenter{\hbox{$\displaystyle 1+ \cfrac{1}{\left(1+\cfrac{\beta^{\prime}r_b^2}{4}\right)^2}      $}}\right]. \notag \\
\end{align*}

This is identical to Eq.~(46) in Ref.~\citenum{lu2006nonlinearphysplasma} except $\beta$ is replaced by $\beta^{\prime}$. It is worth recalling  that an underlying assumption of Eq.~(\ref{eq:rbeq}) is that the $r_b$ dependence in $\beta^{\prime}$ arises from $h$, $c_1$ and $c_2$. 

Once $r_b(\xi)$ is calculated by integrating Eq.~(\ref{eq:rbeq}) starting from $r_m$, the wake potential described by Eq.~(\ref{eq:psi_final}) can be used to obtain the longitudinal electric field in the back half of the bubble $(r \leq r_b(\xi))$ 

\begin{align}
\label{eq:efield}
E_z(\xi) = \frac{d}{d\xi} \psi_0(\xi) = D^{\prime}(r_b) r_b \frac{dr_b}{d\xi}
\end{align}
where $D^{\prime}(r_b) \equiv \frac{1}{2} + \frac{\beta^{\prime}}{2} + \frac{1}{4}r_b \frac{d\beta^{\prime}}{dr_b}$. The slope of the electric field follows directly from Eq.~(\ref{eq:efield})

\begin{align}
\label{eq:efield_prime}
\frac{dE_z}{d\xi} = \ D^{\prime}(r_b) r_b \frac{d^2r_b}{d\xi^2} + F^{\prime}(r_b) \left(\frac{dr_b}{d\xi}\right)^2
\end{align}
where $F^{\prime}(r_b) \equiv D^{\prime}(r_b) + \frac{3}{4}r_b\frac{d\beta^{\prime}}{dr_b} + \frac{1}{4}r_b^2 \frac{d^2 \beta^{\prime}}{dr_b^2}$.

 \begin{figure}[t]
\includegraphics[width=0.5\textwidth]{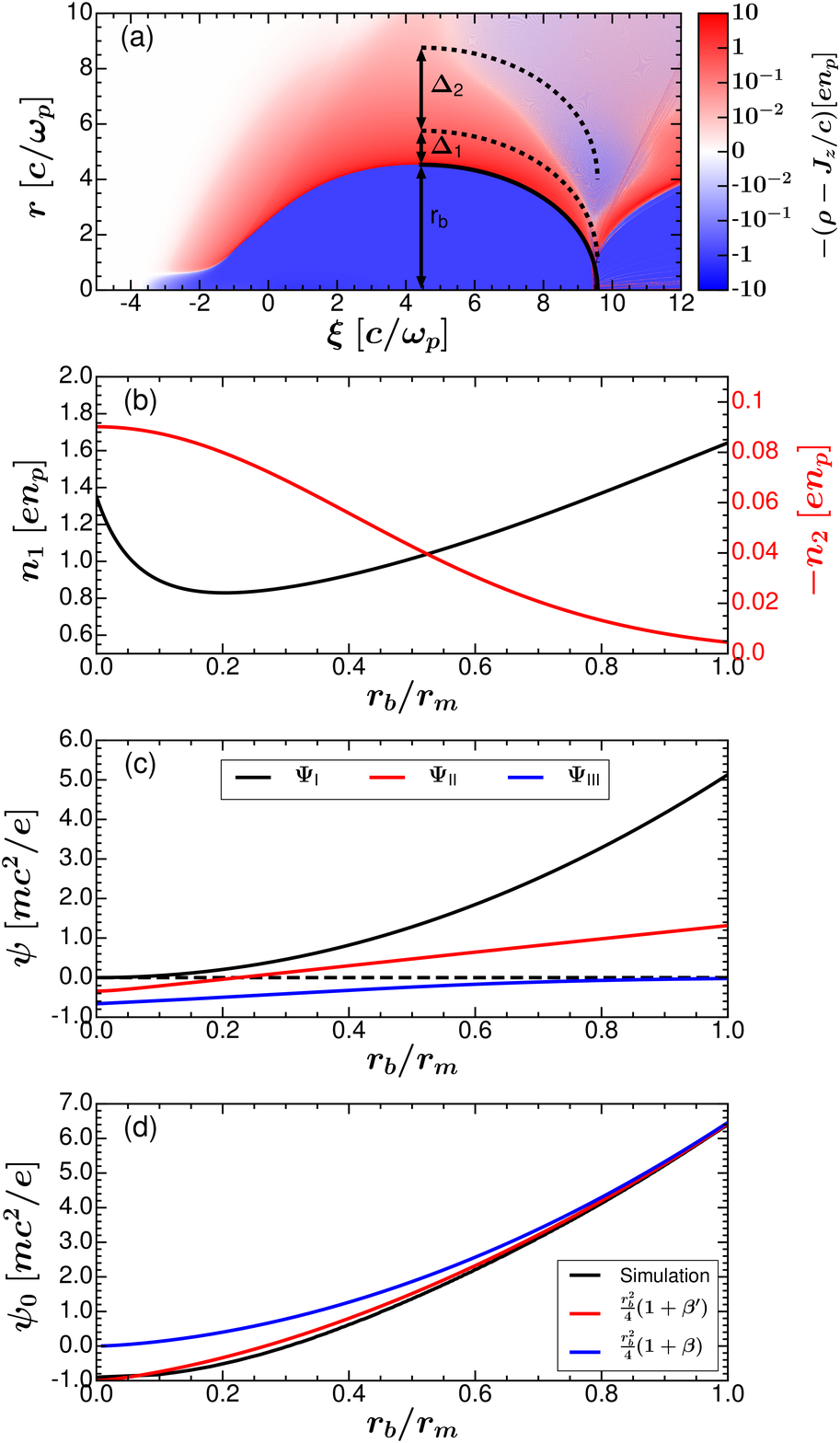}
\caption{\label{fig:example} (a) Contour plot of source term $S$ for a driver with $\Lambda_d = 6$, $\gamma_b = 20000$, $k_p\sigma_z = 1$, and $k_p \xi_c = 0$. The maximum bubble radius is $k_pr_m \simeq 4.53$. The bubble radius $r_b$ [Eq.~(\ref{eq:rbeq})] is calculated using the multi-sheath model $\beta^{\prime}$ with sheath widths $\Delta_1= \Delta_{10} + \Delta_s = 1+0.05r_b$ and $\Delta_2 = \Delta_{20}  = 3$ annotated in black. We use $n_2 = n_{20} e^{-s r_b^2/r_m^2}$ where $s=3$ and $n_{20}$ is calculated from Eq.~(\ref{eq:n2_limit}) using $\psi_{min} = -1$. (b)  $n_1$ [Eq.~(\ref{eq:constant1_eval2})] (black) and $n_2$ (red) , (c) $\Psi_\text{I}$ (black), $\Psi_\text{II}$ (red), $\Psi_\text{III}$ (blue) from Eq.~(\ref{eq:psi_integral}), and (d) $\psi_0$ [Eq.~(\ref{eq:psi_final})] (red) and simulation data (black) plotted as a function of $r_b/r_m$.  The wake potential from single-sheath model~\cite{lu2006nonlinearphysplasma,lu2006nonlineartheoryprl} (blue) is obtained using Eq.~(\ref{eq:psi_final}) with $n_{20} = 0$ [Eq.~(\ref{eq:n2_limit})] and $\psi_{min} = 0$.    }
\end{figure}

In Fig.~\ref{fig:example}(a), we plot the bubble radius $r_b(\xi)$ numerically obtained by integrating Eq.~(\ref{eq:rbeq}) and the sheath widths $\Delta_1(r_b)$ and $\Delta_2(r_b)$ on top of the actual source term $S$ from the simulation shown in Fig.~\ref{fig:schematic1} using $\Delta_{10} = 1$, $\Delta_s = \epsilon r_b = 0.05r_b$, and $\Delta_{20} = 3$. The maximum bubble radius is $r_m \simeq 4.53$. We also plot $n_1(r_b)$ and $n_2(r_b)$ where the $n_{20}$ parameter is calculated from Eq.~(\ref{eq:n2_limit}) using $\psi_{min} = -1$ and $n_1$ is calculated from Eq.~(\ref{eq:constant1_eval2}). Excellent agreement is observed in the bubble trajectory $r_b(\xi)$ calculated using the multi-sheath model and simulation results as seen in Fig.~\ref{fig:example}(a).  

It can be seen in Fig.~\ref{fig:example}(a) that the model for $\Delta_1$ captures the most important regions of the innermost plasma sheath $(S > 0)$ along the bubble interface. While the constant width profile for $\Delta_2$ sufficiently characterizes the $S<0$ region at the rear of the wake, it does not precisely track the empirical second sheath width at the top of the bubble.  However, as shown in Fig.~\ref{fig:example}(b), the profile of $n_2$ (red) decays exponentially to zero near $r_m$ and, therefore, the exact profile of $\Delta_2$ is irrelevant in this region. Near the top of the bubble, $n_1$ (black) can also be well-approximated by $\frac{1}{(1+\alpha_1)^2 -1}$ because $n_2$ approaches zero. When $r_b$ goes to $0$, $n_1$ remains finite because of the negative $n_2$ term in the continuity equation [Eq.~(\ref{eq:constant1_eval2})], which is consistent with the physical picture described earlier and shown in Fig.~\ref{fig:example}(a). However, in the single-sheath model, $n_1 = n_{\Delta} \rightarrow 0$ for $r_b \rightarrow 0$.

The limiting contributions from each of the three regions can also be characterized by the respective wake potential terms defined in Eq.~(\ref{eq:psi_integral}) and plotted in Fig.~\ref{fig:example}(c). When $r_b$ approaches $r_m$, the ion term $\Psi_{\text{I}}$ (black) clearly dominates, the sheath term $\Psi_{\text{II}}$ (red) is in on the order of unity, and $\Psi_{\text{III}}$ (blue) can be neglected because $n_2$ goes to zero. However, when $r_b/r_m \ll 1$, the order of importance is reversed, where $\Psi_{\text{III}}$ is now the most negative component, $\Psi_{\text{II}}$ is less negative, and $\Psi_{\text{I}}$ approaches zero since $r_b \rightarrow 0$. 

Whereas in Refs.~\citenum{lu2006nonlinearphysplasma} and \citenum{lu2006nonlineartheoryprl} the sheath potential term $\Psi_{\text{II}}$ was modeled as positive definite, it can now flip sign because $n_1$ remains finite at the rear of the wake instead of going to zero. Combining all three terms, we observe strong agreement between $\psi_0$ calculated from Eq.~(\ref{eq:psi_integral}) (red) and the on-axis wake potential obtained from the simulation data (black) in Fig.~\ref{fig:example}(d). It is also worth noting that we can recover the single-sheath model, which assumed that $S \geq 0$ outside of the ion bubble $r > r_b(\xi)$, by setting $n_2 =0$ everywhere. Under this assumption, $n_1$ becomes $n_{\Delta} = \frac{1}{(1+\alpha_1)^2-1}$ and $\beta^{\prime}$ becomes $\beta = \frac{(1+\alpha_1)^2 \ln(1+\alpha_1)^2}{(1+\alpha_1)^2-1}-1$ \cite{lu2006nonlinearphysplasma,lu2006nonlineartheoryprl}.  It is clear from Eq.~(\ref{eq:n2_limit}) that $\psi_{min} = 0$ for such a model. We plot $\psi_0$ (blue) in Fig.~\ref{fig:example}(d) obtained by reintegrating Eq.~(\ref{eq:rbeq}) to obtain $r_b$ with $n_{20} = 0$ and all other parameters kept the same. It is clear that the result from the single-sheath model begins to deviate from the simulation results for $r_b/r_m \lesssim 0.7$. This shows that although the single-sheath model is reasonable for such values, $\psi_0$ still deviates because it is connecting to an incorrect value for $r_b \rightarrow 0$.

\begin{figure}[t]
\includegraphics[width=0.5\textwidth]{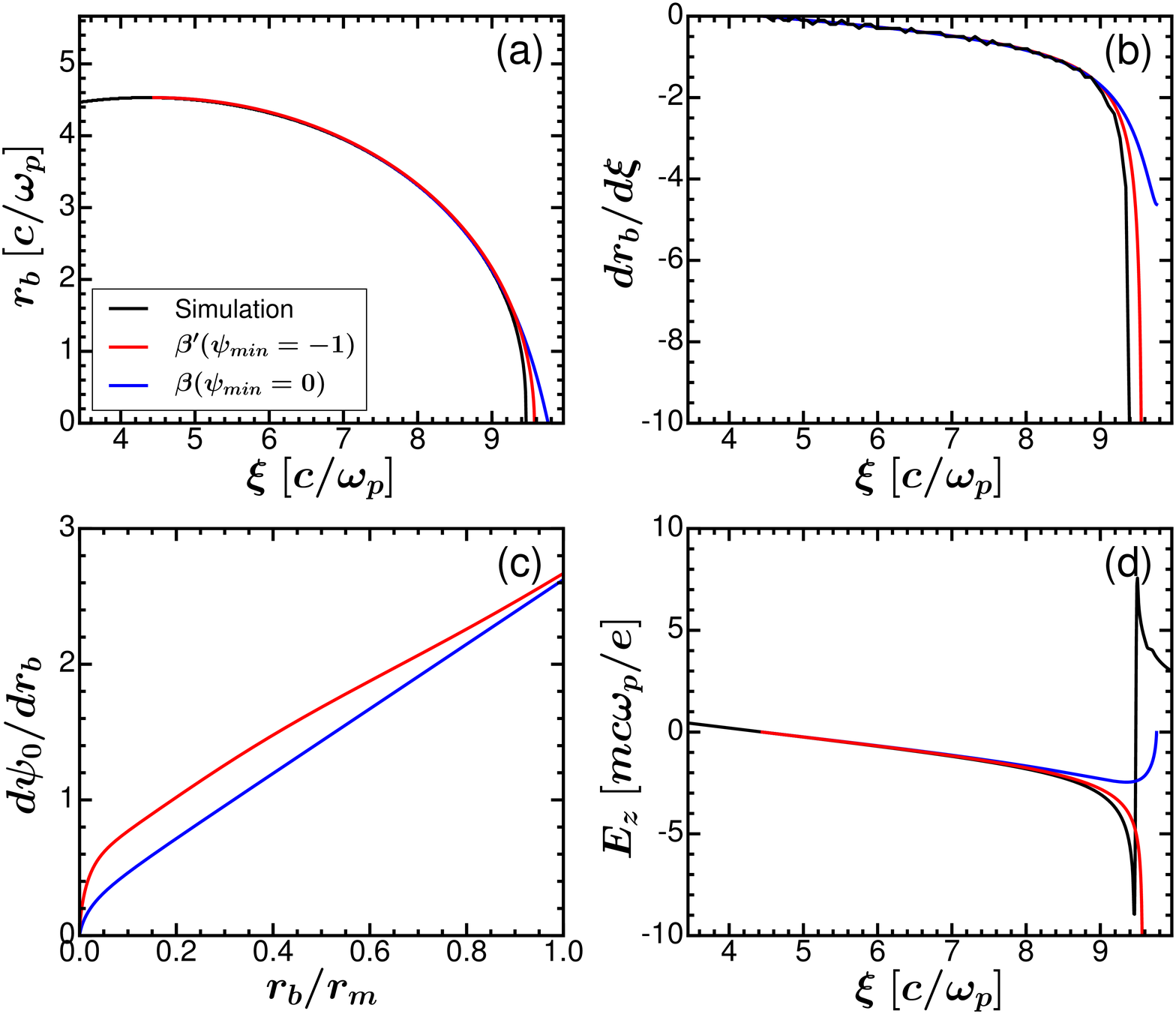}
\caption{\label{fig:example_compare_analysis} Comparisons of (a) $r_b$, (b) $\frac{dr_b}{d\xi}$, (c) $\frac{d\psi_0}{dr_b}$ and (d) $E_z$ obtained from PIC simulation results (black) and numerical calculations (red, blue) for the wakefield shown in Fig.~(\ref{fig:example}). Integration parameters are $\Delta_{1} = 1+0.05r_b$, $\Delta_{2} = 3$, and $n_2 = n_{20} e^{-s r_b^2/r_m^2}$, where $s=3$ and $n_{20}$ is calculated from Eq.~(\ref{eq:n2_limit}) using multi-sheath model $\beta^{\prime}$ with $\psi_{min} = -1$ (red) and single-sheath model $\beta$ with $\psi_{min} = 0$ (blue).  }
\end{figure}

In Fig.~\ref{fig:example_compare_analysis}(a), we plot numerical calculations of the bubble trajectory $r_b(\xi)$ using the multi-sheath model $\beta^{\prime}$ with $\psi_{min} = -1$ (red) and single-sheath model $\beta$ with $\psi_{min} = 0$ (blue) along with $r_b(\xi)$ obtained from PIC simulation results (black). It can be readily seen that the addition of a second plasma sheath acts to bend the electron trajectories toward the axis sooner, thus shortening the predicted wavelength. As a result, the multi-sheath model demonstrates improved agreement with the simulation results over the single-sheath model. The progressively more negative slope $\frac{dr_b}{d\xi}$ observed in the simulation results is due to the fact that sheath electrons copropagate with the wake, i.e., $v_z \sim 1$, as they approach the axis where $\psi_{min} \approx -1$. Thus, they exhibit virtually no phase slippage $d \xi = (1-v_z)dt \approx 0$ for a given change in bubble radius $dr_b$ in this region. In fact, we can show mathematically why this also occurs in the multi-sheath model by rewriting term $\frac{dr_b}{d\xi}$. Using the constant of motion $\gamma - P_z  = 1 + \psi$ \cite{mora}, we find that

\begin{align}
\label{eq:drb}
\frac{dr_b}{d\xi} = \frac{dr_b/dt}{d\xi/dt} =  \frac{v_{\perp}}{1-v_z} = \frac{P_{\perp}}{1+\psi}.
\end{align}

For the innermost sheath electrons returning back the axis, the numerator $P_{\perp}$ is known to be negative and finite. However, the denominator will depend on what kind of model is used for $\psi$.  For the single-sheath model wherein $\psi_{min}=0$, the slope of the trajectory is limited by $\frac{dr_b}{d\xi} \approx P_{\perp}$ near the axis. However, for the multi-sheath model which employs $\psi_{min} = -1$, the slope $\frac{dr_b}{d\xi} $ approaches $-\infty$ near the axis where the denominator $(1+\psi)$ approaches zero. 

As shown in Fig.~\ref{fig:example_compare_analysis}(b), the asymptotic behavior of $\frac{dr_b}{d\xi}$ predicted by the multi-sheath model (red) at the rear of the wake is also borne out in the PIC simulation results (black) where the observed minimum wake potential is close to $-1$. This is an important point because the derived expression for the electric field $E_z(\xi)$  [Eq.~(\ref{eq:efield})] not only depends on $r_b$ but also on the slope of the trajectory $\frac{dr_b}{d\xi}$. Since both the multi-sheath and single-sheath models for $\psi(r_b)$ are functions of only $r_b$, we can express the electric field as $E_z = \frac{d\psi_0}{dr_b} \frac{dr_b}{d\xi}$. As can be seen in Fig.~\ref{fig:example_compare_analysis}(c), the slope of the wake potential $\frac{d\psi_0}{dr_b}$ is larger for the multi-sheath model across all $r_b$ than it is for the single-sheath model due to the larger peak to trough amplitude of $\psi_0$ when using $\psi_{min} = -1$ rather than $\psi_{min} = 0$. It is this term $\frac{d\psi_0}{dr_b}$ which is initially responsible for the more negative electric fields obtained using the multi-sheath model in the range $8 \lesssim k_p\xi \lesssim 9$ seen in Fig.~\ref{fig:example}(d) where the slope of the trajectories $\frac{dr_b}{d\xi}$ are largely similar for both models and simulation results. At the rear of the wake where $r_b/r_m \ll 1$, the slope of the potential $\frac{d\psi_0}{dr_b}$ is small and approaches zero near the axis for both single and multi-sheath models. Since $\frac{dr_b}{d\xi}$ is finite for the single-sheath model, the calculated electric field increases to zero at the rear of the wake $k_p\xi \gtrsim 9$ as $\frac{d\psi_0}{dr_b}$ decreases. In contrast, $\frac{dr_b}{d\xi}$ approaches $-\infty$ for the multi-sheath model near the axis resulting in the characteristic negative spike in the electric field at the rear of the wake observed in PIC simulations results. Therefore, for highly nonlinear plasma wakes, the multi-sheath model employing negative $\psi_{min}$ is needed to predict the electric fields at the back of the bubble, which is a region of interest for accelerating self-injected and trailing bunches.

 \begin{figure*}[t]
\includegraphics[width=1\textwidth]{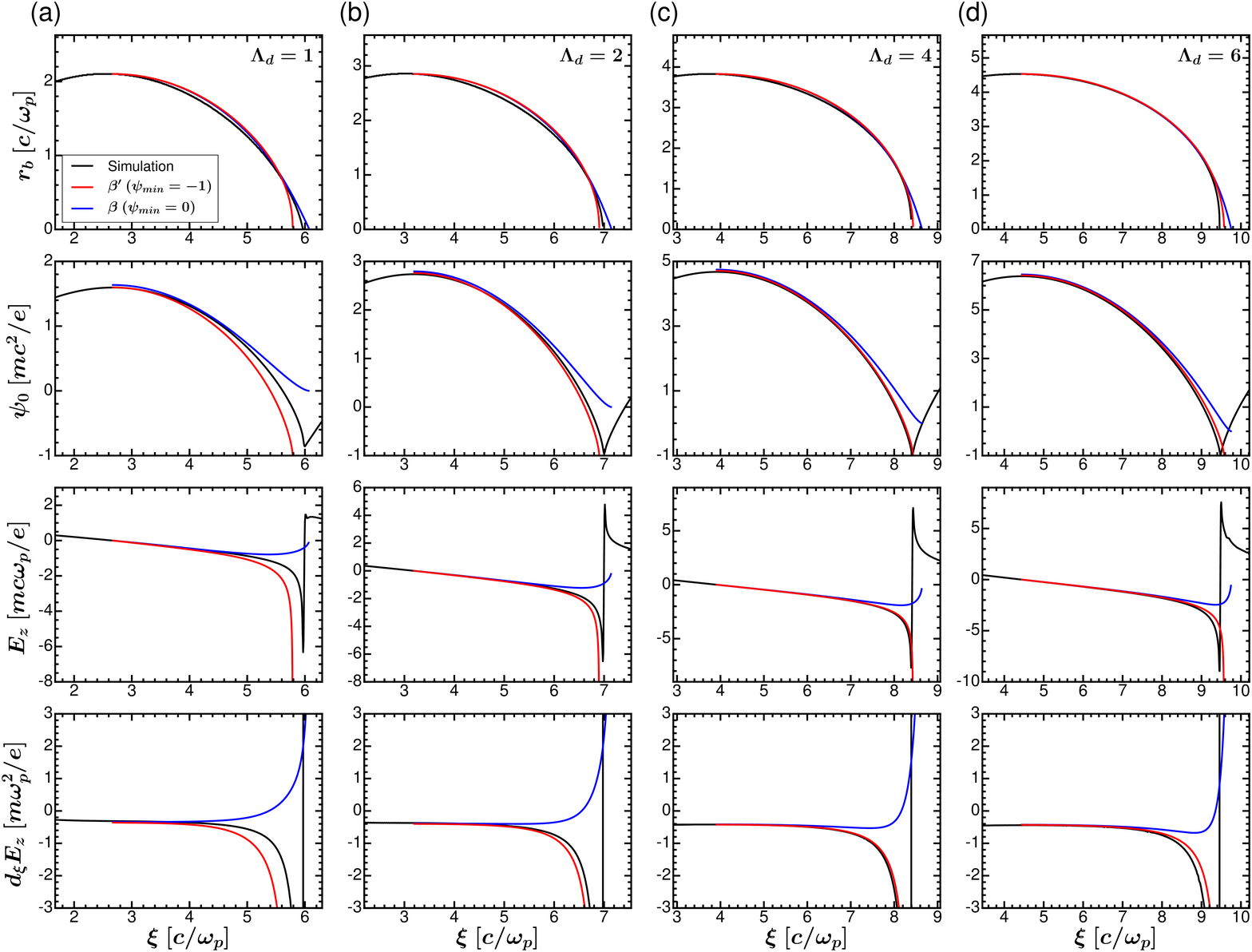}
\caption{\label{fig:unloaded} Comparisons of simulations (black) and numerical calculations (red, blue) of $r_b$, $\psi_0$, $E_z$, and $d_{\xi} E_z$ using Eqs.~(\ref{eq:psi_final})-(\ref{eq:n2_limit}) for plasma wakes excited by bi-Gaussian electron drivers. The peak charge per unit length $\Lambda_d$ is 1, 2, 4 and 6 and the blowout radius $k_pr_m$ is 2.10, 2.85, 3.83 and 4.53 in figures (a)-(d), respectively. Nonevolving electron drivers with energy $\gamma_b=20000$, centroid $k_p\xi_c = 0$, $k_p\sigma_z = 1$, and $k_p\sigma_r = \frac{1}{10}\sqrt{\Lambda_d}$ were used. The integration parameters $ \Delta_{10} =1$, $\Delta_s = 0.05r_b$, $\Delta_{20}=3 $ and $s=3$ are used for all numerical calculations. The multi-sheath model $\beta^{\prime}$ (red) calculates $n_{20}$ [Eq.~(\ref{eq:n2_limit})] using $\psi_{min}= -1$ and single-sheath model $\beta$ \cite{lu2006nonlinearphysplasma,lu2006nonlineartheoryprl} (blue) uses $\psi_{min} = 0$ and $n_2 = 0$ everywhere. $n_1$ is determined from Eq.~(\ref{eq:constant1_eval2}).  }
\end{figure*}

\section{Comparisons of Plasma wakefield Theory and Simulations}
\label{sec:validation}
In the work of Tzoufras et al.~\cite{tzoufrasprab,tzoufrasprl}, it was shown that beam loading in nonlinear plasma wakes can be viewed as a modification to the trajectory of $r_b(\xi)$ due to the presence of a witness beam with a normalized charge per unit length $\lambda(\xi)$. Implicit in such an analysis is the assumption that the theory of Lu et al.~\cite{lu2006nonlinearphysplasma,lu2006nonlineartheoryprl} provides a reasonable prediction for $r_b(\xi)$ (and hence the fields) due to the drive beam. However, the single-sheath model used by Lu et al. does not do as well in the second half of the bubble particularly where a witness beam would be loaded. 

In this section, we examine the predictions of Eqs.~(\ref{eq:rbeq})-(\ref{eq:efield_prime}) for witness beams with specified $\lambda(\xi)$ and compare the numerical results with simulation results obtained using the PIC code {\scshape osiris} \cite{osiris} for various examples of nonlinear plasma wakefields in the blowout regime. We also show how the multi-sheath model improves upon previous results by comparing it to the single-sheath model from Refs.~\citenum{lu2006nonlinearphysplasma} and \citenum{lu2006nonlineartheoryprl}. The purpose of these comparisons is to show that the new multi-sheath model can be used to accurately predict the wake potential $\psi$ and electric field $E_z$ at the rear of an unloaded plasma wake and in a loaded wake with a known trailing bunch profile $\lambda(\xi)$. In a subsequent section, we discuss how to use the multi-sheath model to determine a profile $\lambda(\xi)$ of a witness beam that leads to a desired loaded wakefield $E_z(\xi)$ and compare the results to those in Tzoufras et al.~\cite{tzoufrasprab,tzoufrasprl}.

\subsection{Unloaded Plasma Wakes}

We first examine several cases where an electron drive bunch is used to excite an unloaded plasma wake. As mentioned previously, we are interested in plasma wakes where $r_m \gtrsim 3$. For a bi-Gaussian driver with $k_p\sigma_z \sim 1$, this corresponds to $\Lambda_d \gtrsim 2$ since $r_m \approx 2\sqrt{\Lambda_d}$ \cite{lu2006nonlinearphysplasma,lu2006nonlineartheoryprl}. For these parameters, sheath electrons that trace $r_b(\xi)$ can be accelerated to high velocities $v_z \sim 1$ as they approach the $\xi$-axis where $\psi_{min}$ can be well-approximated by $-1$ \cite{xu2017downrampinj,Thamine2020}. This regime is important because most injection schemes rely on accelerating sheath electrons into the plasma wake at the back of the bubble by temporarily decreasing the phase velocity of the wake. Once injected, these electrons can be accelerated to GeV energies with ultra-high gradients. This region is also interesting because the accelerating fields $E_z$ and transformer ratios are largest for electron bunches at the rear of the wake. However, in order to model the effects of beam loading in this region, we must first be able to capture the behavior of the plasma wake in the absence of any externally injected or trailing bunch.

\begin{figure*}[t]
\includegraphics[width=1\textwidth]{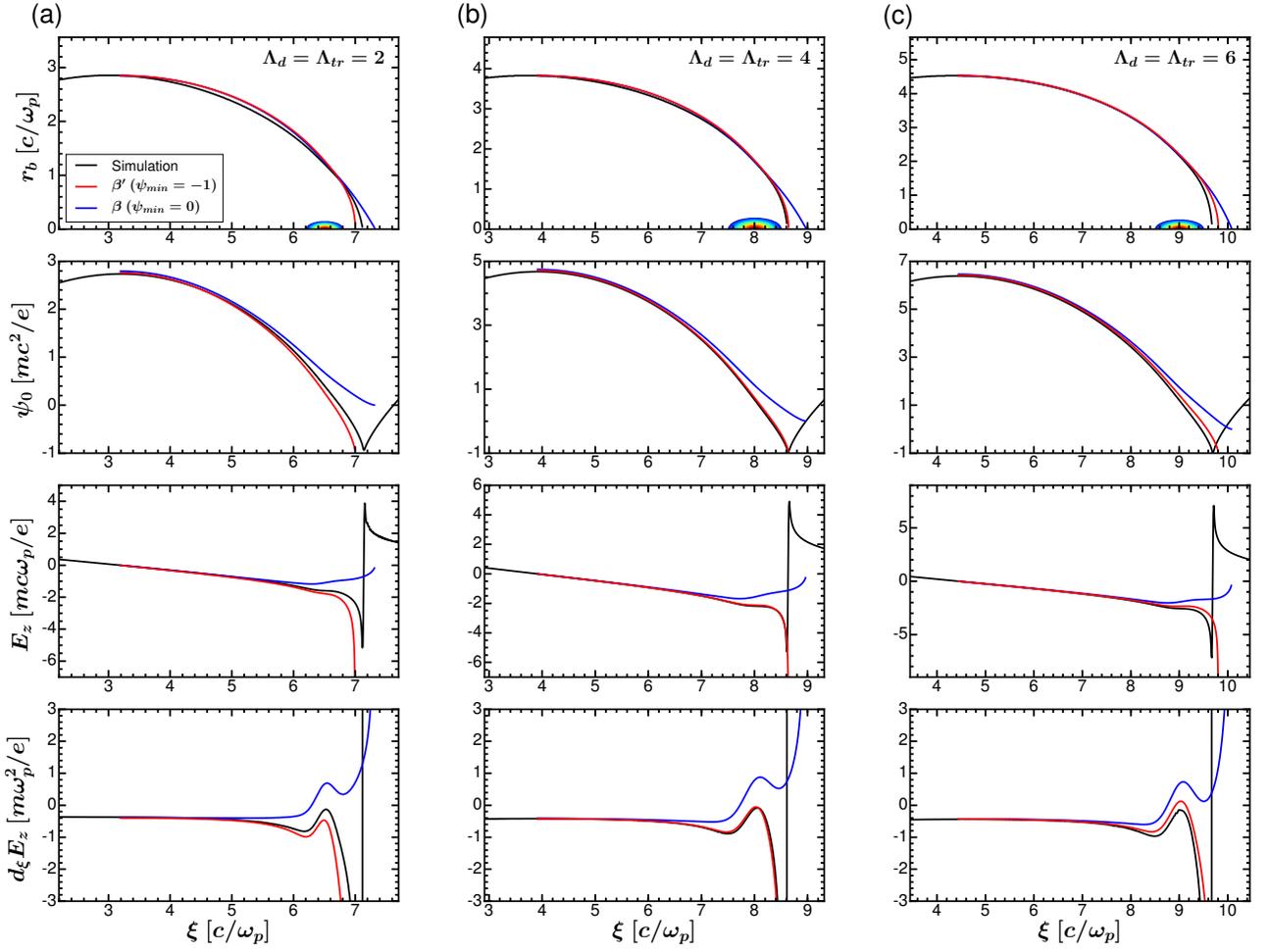}
\caption{\label{fig:gaussian} Comparisons of simulations (black) and numerical calculations (red, blue) of $r_b$, $\psi_0$, $E_z$, and $d_{\xi} E_z$ using Eqs.~(\ref{eq:psi_final})-(\ref{eq:n2_limit}) with bi-Gaussian drive and trailing bunches. The drivers used are identical to those described in Fig.~\ref{fig:unloaded} for each $\Lambda_d$. The trailing bunch parameters are $k_p\xi_{c2} = 6.5,8,9$, $k_p\sigma_{r2} = 0.08, 0.14, 0.14$ and $k_p\sigma_{z2} = 0.15, 0.25, 0.25$ in figures (a)-(c), respectively. All bunches are nonevolving with $\gamma_b = 20000$ and the peak charge per unit length of the driver $\Lambda_d$ and trailing bunch $\Lambda_{tr}$ are the same in each case. The integration parameters $\{ \Delta_{10}, \Delta_s, \Delta_{20} , s\}$ are the same as those used in Fig.~\ref{fig:unloaded}. The multi-sheath model $\beta^{\prime}$ (red) calculates $n_{20}$ [Eq.~(\ref{eq:n2_limit})] using $\psi_{min}= -1$ and the single-sheath model $\beta$ (blue) uses $\psi_{min} = 0$ and $n_2 = 0$ everywhere.  For all calculations, $n_1$ is determined from Eq.~(\ref{eq:constant1_eval2}).  }
\end{figure*}

In Fig.~\ref{fig:unloaded}, we plot the numerical calculations of the bubble trajectory $r_b(\xi)$, potential $\psi_0(\xi)$, electric field $E_z(\xi)$, and electric field slope $d_{\xi}E_z(\xi)$ from Eqs.~(\ref{eq:psi_final})-(\ref{eq:n2_limit}) using the multi-sheath model $\beta^{\prime}$ (red) and single-sheath model $\beta$ (blue) together with the simulations results (black curve) for electron drivers with different $\Lambda_d$ ranging from 1 to 6. Nonevolving drivers were used with $\gamma_b = 20000$, $k_p\sigma_z = 1$, and $k_p \sigma_r = \frac{\sqrt{\Lambda_d}}{10}\approx \frac{r_m}{20}$. The same profiles $\Delta_1 = 1 + 0.05r_b$, $\Delta_2 = 3$, $n_2 = n_{20} e^{-sr_b^2/r_m^2}$ and $s=3$ are used for all calculations. The multi-sheath model calculates $n_{20}$ [Eq.~(\ref{eq:n2_limit})] using $\psi_{min}  = -1$ while single-sheath model uses $n_2 = 0$ everywhere and, therefore, $\psi_{min} = 0$.

In each case, strong agreement is observed between the calculated bubble radius $r_b(\xi)$ and the simulation results along regions where $r_b$ is close to the maximum blowout radius $r_m$ and $n_2$ can be neglected due to its exponential profile. It is only at the rear of the wake that the trajectories $r_b(\xi)$ of the single-sheath and multi-sheath models begin to deviate due to inclusion of the negative source term $n_2 $ which allows for $\psi_{min} = -1$. 

As noted previously, the negative wake potential near the axis employed by the multi-sheath model and observed in PIC simulations is responsible for the bubble trajectories bending back to the axis with large negative slopes $\frac{dr_b}{d\xi}$ from Eq.~(\ref{eq:drb}) and hence large negative values of $E_z(\xi) = \frac{d\psi}{d\xi} $. This leads to the multi-sheath model providing better agreement with the simulation results at the rear of the wake when compared to the single-sheath model. Although not shown for $\Lambda_d =1$, the multi-sheath still works well if a less negative value for $\psi_{min}$ is used. From simulation results, it can seen that $\psi_{min} \approx -0.85$ for $\Lambda_d = 1$. Therefore, from Eq.~(\ref{eq:drb}), the slope of the trajectory $\frac{dr_b}{d\xi}$ from the PIC simulation does not bend as much as the that of the multi-sheath model near the axis. Thus, it is possible to improve the results by tailoring $\psi_{min}$ from PIC simulation data for drivers with $\Lambda_d \lesssim 1$.

By construction, the on-axis wake potentials $\psi_0(\xi)$ differ at the very rear of the wake. Both sheath models predict nearly identical peak potentials $\psi_0(r_m)$, i.e., $\beta(r_m) \approx \beta^{\prime}(r_m)$. However, the values of $\psi_0(r_b)$ differ between the two sheath models for $r_b \lesssim 0.7 r_m$. Since the multi-sheath model covers a larger range of potentials from peak, $\psi_0(r_m)$, to trough, $\psi_{min} \approx -1$, it also exhibits larger $\frac{d\psi_0}{dr_b}$ at all $r_b$ when compared to the single-sheath model. The difference between the two models is more pronounced at lower $\Lambda_d$ since the peak potential scales roughly with the blowout radius squared $\psi_0(r_m) \sim r_m^2$ from Eq.~(\ref{eq:psi_final}) while the minimum wake potentials connect to $\psi_{min} = 0$ for the single-sheath model and $\psi_{min} = -1$ for the multi-sheath model.

In each case, the multi-sheath model produces a monotonically decreasing electric field with a characteristic negative spike near the axis, which is also borne out in PIC simulation results. However, this characteristic spike is absent in the single-sheath model, wherein the electric field actually increases at the rear of the wake in every case. This is also noted in the positive electric field slope $d_{\xi}E_z$ predicted by the single-sheath model at the rear of the wake. By comparison, the multi-sheath model and simulation results indicate that $d_{\xi}E_z$ should remain negative and monotonically decreasing until the innermost electrons reach the $\xi$-axis.

\subsection{Gaussian trailing bunches}
\label{sec:gauss}
We now present several cases in which short bi-Gaussian trailing bunches are placed at the rear of the plasma wakefields shown in Fig.~\ref{fig:unloaded}. The goal is to show that the multi-sheath model provides accurate predictions for beam loading including regions where the wake potential $\psi$ is negative. In Fig.~\ref{fig:gaussian}, we examine several examples in which trailing bunches were added at the back of the same ion channels with centroids located at $k_p\xi_{c2} = 6.5, 8, 9$ and bunch lengths $k_p \sigma_{z2} = 0.15, 0.15, 0.2$. The density profile contours of the narrow bunches are also shown in plots of $r_b(\xi)$ (top row of Fig.~\ref{fig:gaussian}). In each case, non-evolving drive and trailing bunches with the same energy $\gamma_b = 20000$ and peak charge per unit length $\Lambda_{tr} = \Lambda_d$ are used. The multi-sheath model using $\psi_{min}=-1$ is shown in red while the single-sheath model using $\psi_{min} = 0$ is shown in blue. It is clear that the wake potentials $\psi_0(\xi)$ and electric fields $E_z(\xi)$ of the two models diverge at the back of the wake. In every case, the single-sheath model overestimates the electric field in regions where the beam load is present. In contrast, the multi-sheath model accurately captures the behavior of the nearly constant electric field in the center of each beam and exhibits strong agreement with the simulation results. The difference between the two models is also illustrated in plots of the electric field slope $d_{\xi}E_z$.

\subsection{Self-injected bunches}

The multi-sheath model can also be used to characterize the loading of the wake due to self-injection. In this section, we will revisit a recent result published in Ref.~\citenum{Thamine2020} in which a new method of controllable injection was demonstrated using an evolving electron driver. This approach relies on expanding the ion channel by focusing the driver from spots sizes on the order of the blowout radius $r_m$ to spot sizes much less than $r_m$. During this process, the wake velocity $\gamma_{\phi}$ can be significantly reduced and sheath electrons can be injected into the plasma wake near the axis. The driver parameters that control this injection process are the peak current $\Lambda_d$, duration $\sigma_z$, energy $\gamma_b$, and Courant-Snyder (CS) parameters $\beta$, $\alpha$, and $\gamma$ \cite{csparams}, where $\beta = \langle x^2 \rangle/\epsilon$, $\alpha =  -\langle x x^{\prime}  \rangle/\epsilon$, $\gamma = \langle x^{\prime 2} \rangle/\epsilon$, and $\epsilon = \sqrt{\langle x^2 \rangle  \langle x^{\prime 2} \rangle - \langle x x^{\prime}  \rangle^2}$ is the geometric emittance.  For these parameters, the diffraction length of the driver is $\beta^{*} = \sigma_0^2/\epsilon$, where $\sigma_0$ is the focal spot size and the betatron wavenumber is $k_{\beta} = k_p/\sqrt{2 \gamma_b}$. 

\begin{figure}[t]
\includegraphics[width=0.5\textwidth]{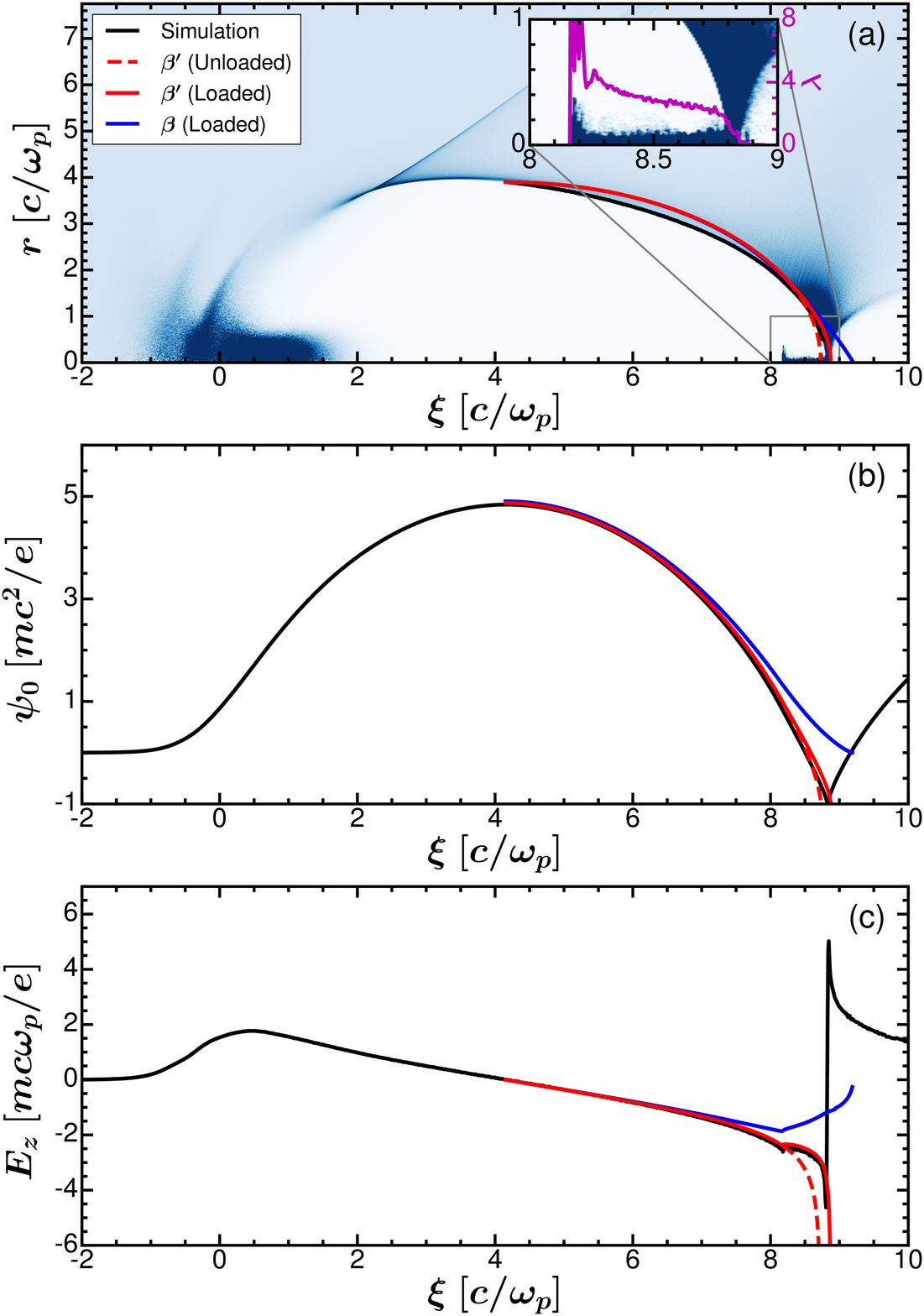}
\caption{\label{fig:injection} (a) Electron density distribution of a plasma wake with a blowout radius $r_m \simeq 3.9$ excited by an evolving electron driver with peak current $\Lambda_d=6$, energy $\gamma_b = 20000$, $k_p \sigma_z =0.7$ after propagating a distance of $1630 ~c/\omega_p$ into a constant density plasma. The driver is initially focused at the plasma entrance with a spot size $k_p\sigma_r =1.225$ and CS parameters $k_p\beta_i \approx 1225$ and $\alpha_i = 0$. The current profile $\lambda(\xi)$ of the injected electrons is shown in purple in the inset plot. Comparisons of simulation results (black) and numerical calculations (dashed red, solid red, solid blue) of (a) $r_b$, (b) $\psi_0$, and (d) $E_z$ using Eqs.~(\ref{eq:psi_final})-(\ref{eq:n2_limit}). For all calculations, the integration parameters $\{ \Delta_{10}=1,\Delta_s=0.05r_b, \Delta_{20}=3 , s=3\}$ are the same as those used in Figs.~\ref{fig:unloaded} and \ref{fig:gaussian}. The dashed and solid red lines correspond to the unloaded and loaded wake calculated using the multi-sheath model $\beta^{\prime}$ with $\psi_{min}= -1$. The blue lines correspond to the loaded wake calculated using the single-sheath model $\beta$ with $\psi_{min} =0$.  }
\end{figure}

In case B from Ref.~\citenum{Thamine2020}, a bi-Gaussian drive bunch with peak current $\Lambda_d=6$, energy $\gamma_b = 20000$, and $k_p \sigma_z =0.7$ is initially focused at the plasma entrance with a spot size of $k_p\sigma_r =1.225$ and CS parameters $k_p \beta_i = k_p\beta^{*} \approx 1225$, and $\alpha_i \approx 0$. Since the driver is not matched, i.e., $k_{\beta}\beta^{*} \approx 6.125$, it is self-focused by the plasma and oscillates at the scale length of the betatron wavelength $2\pi \sqrt{2\gamma_b} c/\omega_p$. The electron density distribution of the plasma wake, driver, and injected beam are shown in Fig.~\ref{fig:injection}(a) after the driver has propagated a distance $k_p z =1630$ into the constant shelf density plasma. The blowout radius at this point is $r_m \simeq 3.9$ and each driver scallop corresponds to a full betatron oscillation \cite{blumenthal42GeV85cm}. During the first betatron period, plasma electrons are injected at the rear of the bubble as the spot size of the driver decreases and the wake expands. As seen in the inset plot of Fig.~\ref{fig:injection}(a), the current profile of the injected bunch varies from $\sim$20 to $\sim$40 kA over the core of the bunch. While the spot size continues to oscillate after the initial injection, the bubble remains fully expanded due to beam loading effects and scalloping of the drive bunch.

In Fig.~\ref{fig:injection}, we compare numerical calculations (dashed red, solid red, and solid blue) of the bubble trajectory $r_b(\xi)$, potential $\psi_0(\xi)$, and electric field $E_z(\xi)$ from Eqs.~(\ref{eq:psi_final})-(\ref{eq:n2_limit}) to simulation results (black). The numerical calculations use $\lambda(\xi)$ of the injected bunch taken from the simulation. The integration parameters $\{\Delta_{10}, \Delta_{20}, s\}$ are identical to those used in Figs.~\ref{fig:unloaded} and~\ref{fig:gaussian}. The solid red lines correspond to the loaded wake calculated using the multi-sheath model with $\psi_{min} = -1$ in Eq.~(\ref{eq:n2_limit}) while the solid blue lines correspond to the loaded wake calculated using the single-sheath model with $\psi_{min} = 0$. For reference, we also plot the unloaded wake (dashed red) obtained from the multi-sheath model to illustrate the effects of beam loading from the self-injected bunch. The multi-sheath model agrees very well with the simulation results. This agreement is significant because it shows that it is now possible to model precisely how injected beams load the plasma wake. And this sets the stage for using the multi-sheath model to accurately predict how to shape the witness beam for desired profiles for $E_z(\xi)$.

\begin{figure}[t]
\includegraphics[width=0.5\textwidth]{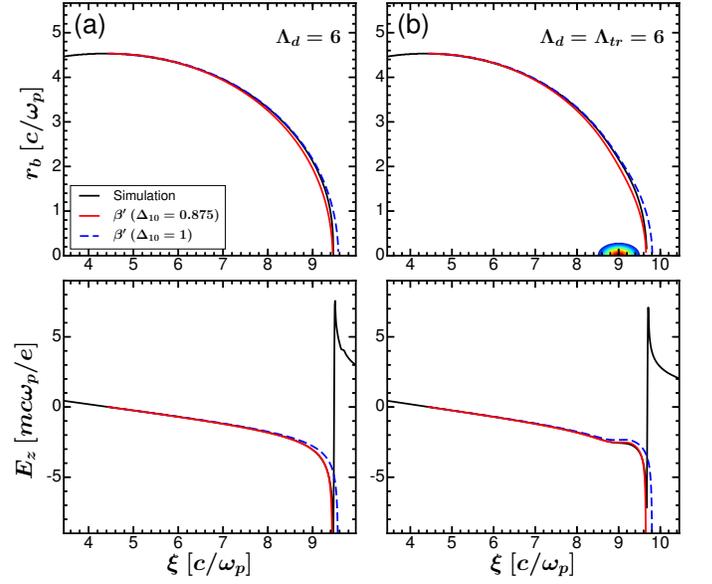}
\caption{\label{fig:optimized} Comparisons of simulations (black) and numerical calculations (red, blue) of $r_b$ (top) and $E_z$ (bottom) using Eqs.~(\ref{eq:psi_final})-(\ref{eq:n2_limit}) for an (a) unloaded and (b) loaded plasma wake. The driver and witness beam parameters are the same as those used for Figs.~\ref{fig:unloaded}(d) and \ref{fig:gaussian}(c). Results are shown using the multi-sheath model $\beta^{\prime}$ with $\Delta_{10}=0.875$ (red) and $\Delta_{10}$ = 1 (dashed blue). The parameters \{$\Delta_s=0.05r_b$, $\Delta_{20} = 3$, $s=3$, and $\psi_{min} = -1$\} are the same as those used in Figs.~\ref{fig:unloaded} and \ref{fig:gaussian}.  }
\end{figure}

\subsection{Phenomenological parameter optimization for beam loading}

In the results presented so far, we have shown that the multi-sheath model reproduces the qualitative plasma wake features in a various cases using a fixed set of phenomenological parameters $\{\Delta_{10}=  1, \Delta_s = 0.05r_b, \Delta_{20} = 3, s=3, \psi_{min} = -1\}$. In some instances, however, there can be a slight mismatch between the wake length predicted by the multi-sheath model and observed in PIC simulations. The underlying reason is that the expression for $\psi$ employed by the multi-sheath model does not perfectly match the empirical wake potential. Therefore, the trajectory [Eq.~(\ref{eq:rbeq})] can slightly undershoot or overshoot the PIC simulation results. 

In Fig.~\ref{fig:optimized}(a), we show how the trajectory obtained using the multi-sheath model can be adjusted by tuning the parameter $\Delta_{10}$ for the plasma wake shown in Fig.~\ref{fig:unloaded}(d). Using $\Delta_{10}=1$ (blue), the multi-sheath model overestimates the the plasma wake length and, therefore, the negative spike in the electric field occurs at a larger $\xi$ when compared to the simulations results. This disagreement can be addressed by reducing the first sheath width $\Delta_{10}$ to decrease the wake length for improved numerical results. By using $\Delta_{10}=0.875$ (solid red), it can be seen that the calculated trajectory now crosses the $\xi$-axis sooner resulting in improved agreement with the simulated bubble length. As a result, the calculated electric field exhibits nearly perfect agreement with the PIC simulation results at the rear of the wake. While $\Delta_{10}$ was lowered to reduce the wake length in this example, it is worth noting that higher values of $\Delta_{10}$ can be used to increase the wake length in other cases. Once the parameter $\Delta_{10}$ is optimized for a particular driver, it can be used for any beam loading calculations involving trailing bunches. In Fig.~\ref{fig:optimized}(b), we show how beam loading results are improved by using the optimized $\Delta_{10}$ with a Gaussian trailing bunch. It can be readily seen that numerical results using $\Delta_{10}=0.875$ provide better agreement with the simulation results for the loaded wakefield and trajectory crossing with the $\xi$-axis.

\section{Designing beam loads for nonlinear plasma wakes}
\label{sec:beamloading}
In the previous section, we showed that the multi-sheath model accurately predicts the bubble trajectories and fields in the second half of unloaded and loaded plasma wakes in the nonlinear blowout regime. We considered situations where the current profile of the trailing bunch $\lambda(\xi)$ was either calculated from the PIC simulation data or specified beforehand. In this section, we show how to design a beam load $\lambda(\xi)$ using Eqs.~(\ref{eq:n2_limit})-(\ref{eq:rbeq}) to produce a specified plasma wakefield $E_z(\xi) = f(\xi)$ for the axial wake potential $\psi_0 = (1+\beta^{\prime})r_b^2/4$. The beam profiles designed using the multi-sheath model will be validated against PIC simulations using {\scshape osiris} \cite{osiris}. Simulation results using the multi-sheath model are compared to results obtained from Tzoufras et al.~\cite{tzoufrasprl, tzoufrasprab}. In the subsequent section, we discuss the differences between this work and Refs.~\citenum{tzoufrasprl} and \citenum{tzoufrasprab}.
\subsection{Exact solutions for loading arbitrary wakefields}
\label{sec:arbwake}
We consider here a general methodology for loading wakefields of arbitrary profiles $E_z( \xi_t \leq \xi \leq \xi_f) = f(\xi)$ when the current profile of the bunch has a well-defined beginning (head) at $\xi = \xi_t$ and end (tail) at $\xi = \xi_f$. The current $\lambda(\xi)$ profile required to produce the specified wakefield $f(\xi)$ can be obtained using a simple two-step process. In the first step, the unloaded bubble trajectory $r_b(\xi)$ is calculated by integrating Eq.~(\ref{eq:rbeq}) starting from $r_b = r_m$. Once $r_b(\xi)$ is obtained, the unloaded wakefield $E_z(\xi)$ is determined from Eq.~(\ref{eq:efield}). In the second step, the order of operations is reversed. Since the desired loaded wakefield $E_z( \xi_t \leq \xi \leq \xi_f) = f(\xi)$ is known, the loaded bubble trajectory $\tilde{r}_b(\xi)$ can be reversed engineered from Eq.~(\ref{eq:efield}) by numerically integrating the following

\begin{align}
\label{eq:efieldloaded}
\frac{d\tilde{r}_b}{d\xi} = \frac{f(\xi)}{D^{\prime}(\tilde{r}_b) \tilde{r}_b } 
\end{align}
from $\tilde{r}_b(\xi_t) = r_b(\xi_t) \equiv r_t$ to either $\tilde{r}_b(\xi_f)$ or $\tilde{r}_b= 0$, whichever comes first. The function $f(\xi)$ is constrained by boundary conditions at $\xi = \xi_t$, which require wakefield continuity $f(\xi_t) = \lim_{\epsilon \rightarrow 0^-} E_z(\xi_t+\epsilon) \equiv -E_t$. Once $\tilde{r}_b(\xi)$ is calculated, the corresponding wake potential $\psi_0(\tilde{r}_b) = (1 + \beta^{\prime}(\tilde{r}_b)) \tilde{r}_b^2/4$ can be determined. Although Eqs.~(\ref{eq:rbeq})-(\ref{eq:efield_prime}) were derived to solve for $r_b(\xi)$ given $\lambda(\xi)$, conversely they can instead be used to solve for $\lambda(\xi)$ for a given trajectory $\tilde{r}_b(\xi)$. Expressing the derivatives of $\tilde{r}_b(\xi)$ in terms of $f(\xi)$ and $\frac{df}{d\xi}$ using Eqs.~(\ref{eq:efield})-(\ref{eq:efield_prime}), we can rewrite Eq.~(\ref{eq:rbeq}) as

\begin{align}
\label{eq:lambda_exact}
\lambda(\xi) =  C^{\prime}\tilde{r}_b^2 +& \left(\frac{B^{\prime}}{D^{\prime2}} -\frac{A^{\prime}F^{\prime} }{D^{\prime3} \tilde{r}_b^2} \right)f(\xi)^2  +\left(\frac{A^{\prime}}{D^{\prime}}\right)\frac{df(\xi)}{d\xi}.
\end{align}
$f(\xi)$ is the desired loaded electric field and $A^{\prime}(\tilde{r}_b)$, $B^{\prime}(\tilde{r}_b)$, $C^{\prime}(\tilde{r}_b)$, $D^{\prime}(\tilde{r}_b)$, and $F^{\prime}(\tilde{r}_b)$ are specified in Sec.~\ref{sec:potential}. Since the left-hand side must be positive definite $(\lambda(\xi) \geq 0)$ for an electron bunch, the wakefield slope $d_{\xi}f$ is naturally constrained by Eq.~(\ref{eq:lambda_exact}). Physically, this means that the slope of the loaded wakefield $f(\xi)$ cannot be more negative than the corresponding slope in the absence of any load. In cases where the slope $d_{\xi}f$ is sufficiently negative, the current profile calculated from Eq.~(\ref{eq:lambda_exact}) would flip sign $(\lambda(\xi) <0)$ which would require positive charge densities, e.g., positrons, along portions of the beam load which cannot be focused. Although we have not discussed the transverse force, we note that for azimuthally symmetric wakes the focusing force remains perfectly linear even for loaded wakefields.

\subsection{Ultrarelativistic blowout regime}

In the ultrarelativistic limit, where the bubble radius is large $r_b \gg 1$, the sheath terms are small, i.e., $\beta^{\prime} \ll 1$ and $\beta^{\prime} r_b^2/4 \gtrsim 1$, relative to the ion term $r_b^2/4 \gg 1$. Therefore, Eqs.~(\ref{eq:efieldloaded}) and (\ref{eq:lambda_exact}) can be approximated by the leading terms of the coefficients $[A^{\prime}(r_b), B^{\prime}(r_b), C^{\prime}(r_b), D^{\prime}(r_b), F^{\prime}(r_b),\beta^{\prime}(r_b)] \rightarrow (r_b^2/4, 1/2, 1/4, 1/2, 1/2,0)$. In this limit, the loaded bubble trajectory can be expressed as   

\begin{align}
\label{eq:efieldloadedsimplified}
\frac{d\tilde{r}_b}{d\xi} &= \frac{f(\xi)}{ \tilde{r}_b/2 } \notag \\
\tilde{r}_b(\xi)^2 &= r_t^2 + 4\int^{\xi}_{\xi_t}f(\xi)d\xi.
\end{align}
Alternatively, the loaded bubble trajectory can also be expressed as a convolution of the loaded wakefield $f(\xi)$ and the heaviside step function, $H(x)$,

\begin{align}
\label{eq:efieldloadedheaviside}
\tilde{r}_b(\xi)^2 &= r_t^2 + 4\int^{\xi}_{-\infty}f(\xi)H(\xi - \xi_t)d\xi.
\end{align}
In the ultrarelativistic limit, the corresponding current profile of the beam load from Eq.~(\ref{eq:lambda_exact}) becomes

\begin{align}
\label{eq:lambda_simplified}
\lambda(\xi) =  \frac{\tilde{r}_b^2}{4} + f(\xi)^2 +\left(\frac{\tilde{r}_b^2}{2}\right)\frac{df(\xi)}{d\xi}. 
\end{align}
In the absence of a beam load $(\lambda(\xi) = 0)$, it can seen from Eq.~(\ref{eq:lambda_simplified}) that the slope of the electric field inside the bubble $df/d\xi = dE_z/d\xi \simeq -1/2$ when the bubble radius is maximum $r_b = r_m$ and $f = E_z \simeq \frac{1}{2} r_b \frac{dr_b}{d\xi} \simeq 0$. As noted in the previous section, the slope of the loaded wakefield is naturally constrained by Eq.~(\ref{eq:lambda_simplified}). For an electron (or anti-proton) beam $\lambda \geq 0$,

\begin{align}
\label{eq:lambda_simplified_constraint_pp}
\frac{\tilde{r}_b^2}{4} + f(\xi)^2 +\left(\frac{\tilde{r}_b^2}{2}\right)\frac{df(\xi)}{d\xi} \geq 0
\end{align}
from which it follows, 
\begin{align}
\label{eq:lambda_simplified_constraint}
\frac{df(\xi)}{d\xi} \geq -\frac{1}{2}\left( 1 + \frac{4f(\xi)^2}{\tilde{r}_b^2} \right).
\end{align}

Eqs.~(\ref{eq:efieldloadedsimplified}) and (\ref{eq:lambda_simplified}) can be solved analytically for many functions $f(\xi)$. However, any such functions $f(\xi)$ must satisfy Eq.~(\ref{eq:lambda_simplified_constraint}) for all $\xi \in [ \xi_t,\xi_f]$ along the beam load and the continuity constraint at the head of the bunch $f(\xi_t) = -E_t$. In Refs.~\citenum{tzoufrasprl} and \citenum{tzoufrasprab}, analytic solutions to Eqs.~(\ref{eq:efieldloadedsimplified}) and (\ref{eq:lambda_simplified}) were derived for a beam load designed to produced a constant wakefield $f(\xi \geq \xi_t) = -E_t$ and $df/d\xi = 0$ extending from the head of the bunch $\xi_t$ all the way to the rear of the wake $\tilde{r}_b(\xi_f) = 0$. Such a wakefield can be used to accelerate a trailing bunch to multi-GeV energies while maintaining the kinds of low energy spreads needed for next-generation linear collider and XFEL applicaitons. 

For a constant wakefield $df/d\xi = 0$, it is trivial to show that Eq.~(\ref{eq:lambda_simplified_constraint}) is always satisfied and the solution to Eq.~(\ref{eq:efieldloadedsimplified}) is a parabola $\tilde{r}_b^2 = r_t^2 - 4E_t(\xi- \xi_t)$. It follows directly from the loaded bubble trajectory that the maximum length of the beam load $\Delta \xi_{tr} \equiv \xi_f - \xi_t = \frac{r_t^2}{4E_t}$ is limited by length of the bubble $\tilde{r}_b(\xi_f) = 0$. Substituting the loaded trajectory $\tilde{r}_b$ into Eq.~(\ref{eq:lambda_simplified_constraint}), the underlying current profile is given by

\begin{align}
\label{eq:lambdaopt_1}
\lambda(\xi) = E_t^2 + \frac{\tilde{r}_b(\xi)^2}{4} = E_t^2 + \frac{r_t^2}{4} -E_t (\xi - \xi_t).
\end{align}

In Refs.~\citenum{tzoufrasprl} and \citenum{tzoufrasprab}, it was shown that this trapezoidal current profile could be written as 
\begin{align}
\label{eq:lambdaopt}
\lambda(\xi) = \sqrt{E_t^4 + \frac{r_m^4}{2^4}} -E_t (\xi - \xi_t)
\end{align}
by solving for $r_t^2/4 = \sqrt{E_t^4 + r_m^4/2^4} - E_t^2$ in terms of $E_t$ and $r_m$ in the ultrarelativistic limit. In the following sections, we will compare the analytic result [Eq.~(\ref{eq:lambdaopt})] derived by Tzoufras et al.~\cite{tzoufrasprl, tzoufrasprab} in the ultrarelavistic limit $(\beta,\beta^{\prime} \rightarrow 0,0)$ with the exact beam profiles obtained by numerically integrating Eq.~(\ref{eq:efieldloaded}) for both the multi-sheath $\beta^{\prime}$ and single-sheath $\beta$ models.

\subsection{Comparisons of theory and simulation results for loading constant wakefields}

We next use the methodology outlined above to design beam loads that produce constant electric fields extending to the very rear of the bubble. Exact profiles will be calculated numerically from Eqs.~(\ref{eq:efieldloaded}) and (\ref{eq:lambda_exact}) for the multi-sheath $\beta^{\prime}$ and single-sheath models $\beta$. We also present results for beam loads calculating using the analytic theory [Eq.~(\ref{eq:lambdaopt})] in the relativistic limit $(\beta,\beta^{\prime} \rightarrow 0,0)$. Beam profiles obtained for each model will be simulated using the PIC code {\scshape osiris} \cite{osiris}. Finally, we will present examples of loading longitudinally varying electric fields using the multi-sheath model and compare the numerical results to PIC simulations. 

We use $\Delta_{20}= 3$, $n_2 = n_{20} e^{-s r_b^2/r_m^2}$, and $s = 3$ when numerically integrating Eqs.~(\ref{eq:efieldloaded}) and (\ref{eq:lambda_exact}). For each case, $\Delta_{10}$ is first optimized for the multi-sheath model using the unloaded plasma wake. For the multi-sheath model, $n_{20}$ is determined from Eq.~(\ref{eq:n2_limit}), which depends on the $\psi_{min}$ used. For the single-sheath model, $n_{20} = 0$ and, therefore, $n_2 = 0$ everywhere. Reasonable estimates can be obtained for a large parameter space if $\Delta_{10} = 1$.
\begin{figure}[t]
\includegraphics[width=0.5\textwidth]{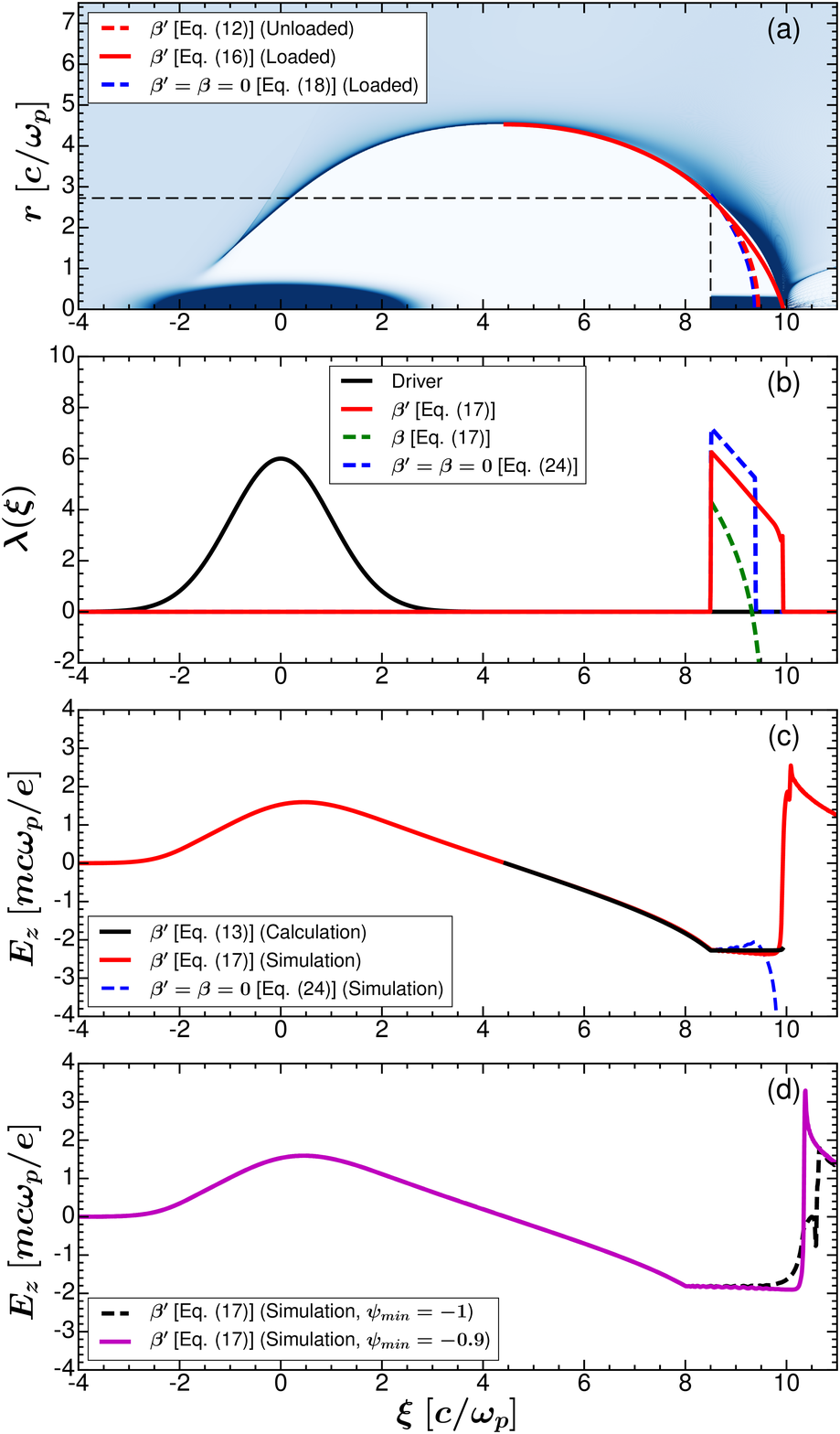}
\caption{\label{fig:beam_loading} (a) Plasma wake excited by a driver $\{\Lambda_d = 6, k_p\sigma_r = 0.245, k_p\sigma_z = 1, k_p\xi_c = 0\}$ with a load [Eq.~(\ref{eq:lambda_exact})] placed at $\xi_t = 8.5$ designed to produce a constant wakefield $E_t = 2.26$ using the multi-sheath model $\beta^{\prime}$ with $\psi_{min} = -1$. The unloaded (loaded) trajectory is shown in dashed (solid) red. (b) Predicted current profiles $\lambda(\xi)$ for the multi-sheath model $\beta^{\prime}$ (red), single-sheath model $\beta$ (green) with $\psi_{min} = 0$, and analytic theory $\beta^{\prime} = \beta = 0$ (blue). (c) Simulated $E_z(\xi)$ using the profiles shown in (b). $E_z(\xi)$ calculated from Eq.~(\ref{eq:efield}) using the $\beta^{\prime}$ current profile is in black. (d) Simulated $E_z(\xi)$ using multi-sheath model $\beta^{\prime}$ [Eq.~(\ref{eq:lambda_exact})] to load the wake at $\xi_t = 8,\ E_t = 1.81$ with $\psi_{min}=$ -0.9 (purple) and -1 (black).}
\end{figure}

In Fig.~\ref{fig:beam_loading}, we compare results for an electron drive bunch with parameters $\Lambda_d=6$, $\gamma_b = 20000$, $k_p \sigma_r = 0.245$, $k_p \sigma_z = 1$ and $k_p \xi_c = 0$. The maximum blowout radius is $r_m \simeq 4.53$. For this driver, we found that a value of $\Delta_{10} = 0.875$ gave the best results (see Fig.~\ref{fig:optimized}). We use $\psi_{min} = -1$ to determine $n_{20}$. The unloaded bubble trajectory $r_b(\xi)$ (dashed red) obtained from integrating Eq.~(\ref{eq:rbeq}) is also shown in Fig.~\ref{fig:beam_loading}(a). We are interested in calculating the current profile of the beam load that can produce a constant wakefield starting at $\xi_t = 8.5$ using the multi-sheath model.  To self-consistently solve for the loaded bubble trajectory $\tilde{r}_b$ using the multi-sheath model, we numerically integrate Eq.~(\ref{eq:efieldloaded}) using $r_t \simeq 2.73$ and $E_t \simeq 2.26$ from the unloaded calculations. The loaded trajectory $\tilde{r}_b$ obtained using the multi-sheath model is plotted (solid red) in Fig.~\ref{fig:beam_loading}(a) and exhibits strong agreement with the simulated wake trajectory produced by the underlying trailing bunch. The loaded trajectory in the ultrarelativistic limit $\tilde{r}_b(\xi)^2 = r_t^2 - 4E_t (\xi- \xi_t)$ is also plotted (dashed blue) using $r_t \simeq 2.81 $ and $E_t \simeq 2.26$ from the PIC simulation data. It can be seen that the loaded parabolic trajectory (analytic result for ultra-relativistic regions of $r_b$) underestimates the length of the plasma wake as it crosses the axis much sooner than expected when compared to the loaded multi-sheath trajectory.

The current profile $\lambda(\xi)$ predicted by the multi-sheath model [Eq.~(\ref{eq:lambda_exact})] is plotted (solid red) in Fig.~\ref{fig:beam_loading}(b). For comparison, we also plot the current profile $\lambda(\xi) \simeq 7.22 - 2.26 \times (\xi- 8.5)$ [Eq.~(\ref{eq:lambdaopt})] obtained in the ultrarelativistic limit (dashed blue line). While the profiles are both trapezoidal, the multi-sheath model predicts a bunch length $\Delta \xi_{tr} \simeq 1.44$ which is $\sim 63\%$ longer that that of the analytic beam loading theory $\Delta \xi_{tr} =\frac{r_t^2}{4E_t} \simeq 0.88$. On the other hand, the analytic theory predicts larger currents along the load compared to the multi-sheath model. Integrating the current profiles, the total loaded charge predicted by the multi-sheath model $Q_{tr}/en_pk_p^{-3} \simeq 41.8$ (solid red) is $\sim 21\%$ higher than the loaded charge predicted by the analytic beam loading theory $Q_{tr}/en_pk_p^{-3} \simeq 34.5$ (dashed blue).

We also plot the beam profile calculated using the single-sheath model (dashed green) by integrating Eqs.~(\ref{eq:efieldloaded})-(\ref{eq:lambda_exact}) with $n_{2} = 0$. For the single-sheath results, we self-consistently sample the bubble radius $r_t \simeq 2.7$ and electric field $E_t \simeq 2.05$ at $\xi_t = 8.5$ from the unloaded wake trajectory integrated from Eq.~(\ref{eq:rbeq}) using $\beta$ instead of $\beta^{\prime}$. Compared to the other profiles, the single-sheath model significantly underestimates the slice currents of the beam load at all longitudinal positions. The disagreement is largely attributed to the fact that the unloaded electric field predicted by the single-sheath model does not capture the characteristic negative spike observed in the multi-sheath model and simulation results in Figs.~\ref{fig:example_compare_analysis} and \ref{fig:unloaded}.  Instead, the slope of the unloaded electric field $d_{\xi}E_z$ predicted by the single-sheath model flips signs from negative to positive at the rear of the wake. As a result, the current profile predicted by the single-sheath model also flips sign $(\lambda <  0)$ at the rear of the wake which corresponds to positive charge densities, i.e. positrons, along portions of the beam load.

In Fig.~\ref{fig:beam_loading}(c), we show the electric fields from PIC simulations using the currents profiles predicted by the multi-sheath model and analytic theory shown in Fig.~\ref{fig:beam_loading}(b). For reference, we also plot the expected electric field from the multi-sheath model using Eq.~(\ref{eq:efield}). The simulation results clearly show that the profile delineated by Eq.~(\ref{eq:lambdaopt}) (dashed blue) does not extend to the very rear of the wake. The simulation results using the multi-sheath model (black) produces a nearly constant electric field $E_z$ over almost the entire length of the load extending all the way to the back of the bubble. The electric field in the simulation is in agreement with the expected field calculated using the multi-sheath model shown in red. The deviation between these two curves is small (the red curve has a slight negative slope) and can be attributed to the minimum wake potential $\psi_{min}$ not being exactly equal to $-1$.  

In cases with longer loads $\Delta \xi_{tr} \gtrsim r_m/2$, higher values of $\psi_{min}$ may be needed to correctly load the wake since the loads, themselves, can modify the electron momenta at the back of the bubble and, therefore, alter the wake potential described by the constant of motion $1+\psi = \bar{\gamma} - P_z$ \cite{mora}. In Fig.~\ref{fig:beam_loading}(d), we show electric field simulation results using beam loads [Eq.~(\ref{eq:lambda_exact})] designed to produce a constant wakefield $E_t \simeq 1.81$ starting at $\xi_t = 8$. The bubble radius is $r_t \simeq 3.31$ at $\xi_t$ from the unloaded PIC simulation. Two different cases are presented for trailing beams where $n_{20}$ is determined from Eq.~(\ref{eq:n2_limit}) using $\psi_{min}= -0.9$ (solid purple) and $\psi_{min} = -1$ (dashed black). While the black curve increases at the back of the bubble $(9.5<\xi<10.5)$, the purple curve remains flat over virtually the entire beam load.  The difference can be attributed to the fact that the underlying beam profile used to load the wakefield in black is calculated using $\psi_{min}=-1$ which is more negative than the empirical value of $\psi_{min} \simeq -0.65$ observed in the PIC simulation with the beam load as predicted from the model. As a result, the current profile overestimates the length $\Delta \xi_{tr}$ over which the wake can be loaded as well as the ion channel radius $r_b$, thus leading to larger currents from Eq.~(\ref{eq:lambda_exact}). 

This issue can be addressed by incrementally increasing the value of $\psi_{min}$ used by the multi-sheath model until it matches the empirical $\psi_{min}$ from the PIC simulation results with the beam load. For $\psi_{min} = -0.9$, we see that the plasma wakefield is nearly perfect loaded over a reduced bunch length in purple. The value $\psi_{min} = -0.9$ is now in good agreement with the minimum wake potential $\psi_{min} \simeq -0.87$ found in the PIC simulation results in purple. Integrating the current profiles of the underlying beam loads, the total loaded charge $Q_{tr}/en_pk_p^{-3} \simeq 55.4$ obtained using the multi-sheath model with $\psi_{min} =-0.9$ is only marginally lower ($\sim 7.5\%$) than the loaded charge $Q_{tr}/en_pk_p^{-3} \simeq 59.9 $ calculated when using $\psi_{min} = -1$.

As in the previous example, the bunch length $\Delta \xi_{tr} \simeq 2.31$ predicted by the multi-sheath model using $\psi_{min} = -0.9$ is $\sim 53 \%$ longer than the optimal bunch length predicted by analytic beam loading theory $\Delta \xi_{tr} =\frac{r_t^2}{4E_t} \simeq 1.51$. As a result, the loaded charge $Q_{tr}/en_pk_p^{-3} \simeq 55.4$ predicted by the multi-sheath model is $\sim 23\%$ higher than the loaded charge $Q_{tr}/en_pk_p^{-3} \simeq 45$ predicted by integrating Eq.~(\ref{eq:lambdaopt}) of the analytic theory. As in the previous example shown in Fig.~\ref{fig:beam_loading}(b), the gain in the loaded charge $Q_{tr}$ is primarily driven by the longer bunch length $\Delta \xi_{tr}$. The increase in $Q_{tr}$ is lower than that of $\Delta \xi_{tr}$ because the slice currents are also lower.

\begin{figure}[t]
\includegraphics[width=0.5\textwidth]{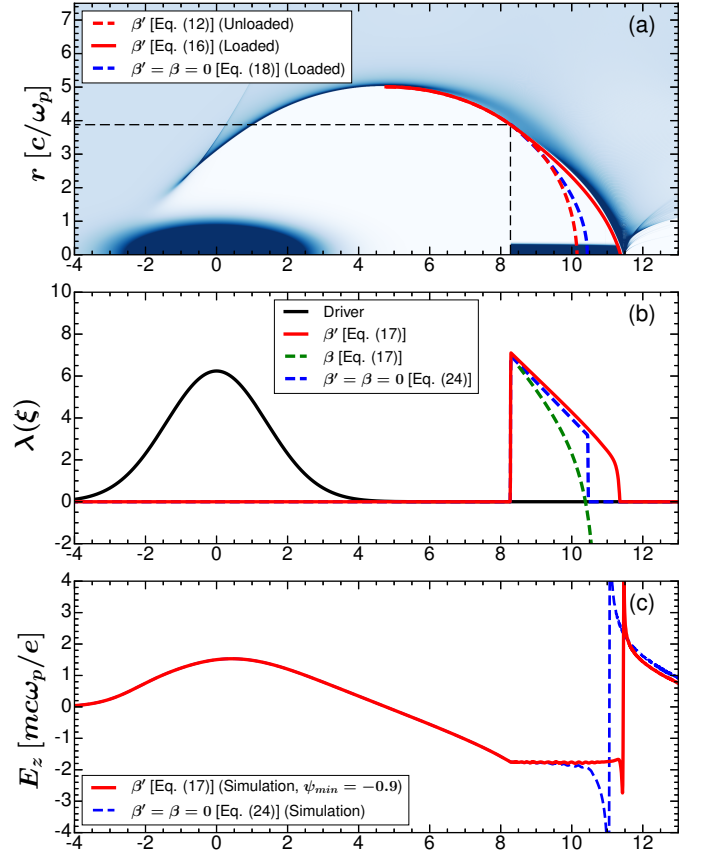}
\caption{\label{fig:beam_loading_tz} (a) Electron density distribution of a plasma wake excited by an electron driver $\{ \Lambda= 6.24, k_p\sigma_r = 0.5, k_p\sigma_z = 1.41, k_p\xi_c =0 \}$ with a load [Eq.~(\ref{eq:lambda_exact})] placed at $\xi_t = 8.27$ designed to produce a constant wakefield $E_t = 1.75$ using the multi-sheath model $\beta^{\prime}$. Integration parameters are $\Delta_{10} = 0.825$, $\Delta_s = 0.05r_b$, $\Delta_{20} = 3$, and $s=3$. The unloaded (dashed red) and loaded (solid red) trajectories are calculated using $\psi_{min} = -1$ and $\psi_{min} = -0.9$, respectively. (b) Current profiles $\lambda(\xi)$ for the multi-sheath model $\beta^{\prime}$ (red) with $\psi_{min} = -0.9$, single-sheath model $\beta$ (green) with $\psi_{min} = 0$, and analytic theory $\beta^{\prime} = \beta = 0$ (blue). (c) Simulated $E_z(\xi)$ using the profiles shown in (b). }
\end{figure}

In the next example, we revisit a beam loading result published by Tzoufras et al.~\cite{tzoufrasprab} in which a bi-Gaussian driver \{$n_b(r,\xi) = [N_b/(2 \pi)^{3/2} \sigma_r^2 \sigma_z] e^{-r^2/(2\sigma_r^2)}e^{-\xi^2/(2\sigma_z^2)}$, $k_p \sigma_r = 0.5$, $k_p\sigma_z = 1.41$, $N_b = 139(c/\omega_p)^3$\} is used to excite a plasma wakefield with a blowout radius $r_m \simeq 5$ in Fig.~\ref{fig:beam_loading_tz}(a). For these parameters, the peak charge per unit length of the driver is $\Lambda_d \simeq 6.24$. 

Using an optimized sheath width $\Delta_{10} = 0.825$ and $\psi_{min} = -1$, the unloaded bubble trajectory (dashed red) is calculated using the multi-sheath model [Eq.~(\ref{eq:rbeq})] and plotted in Fig.~\ref{fig:beam_loading_tz}(a). We are interested in generating a beam profile that can load a constant wakefield $E_t \simeq 1.75$, as previously done in Ref.~\citenum{tzoufrasprab}. From the unloaded PIC simulation results, this electric field occurs at $\xi_t \simeq 8.27$ where the simulated bubble radius is $r_t \simeq 3.91$. Using the multi-sheath model, we self-consistently solve for the loaded bubble trajectory $\tilde{r}_b$ (solid red) by integrating Eq.~(\ref{eq:efieldloaded}) starting at $\xi_t \simeq 8.27$. We use $\psi_{min} = -0.9$ in this example since the bunch length is long, i.e., $\Delta \xi_{tr} \gtrsim r_m/2$. Strong agreement is observed between the loaded trajectory calculated from the multi-sheath model and the trajectory from the PIC simulation with the underlying bunch. For reference, we also plot the loaded trajectory in the ultrarelativistic limit $\tilde{r}_b^2 = r_t^2 - 4E_t(\xi- \xi_t)$ using $r_t$ and $E_t$ from PIC simulation data. Like in the previous example, the parabolic trajectory underestimates the length of the wake the length when compared to the loaded trajectory of the multi-sheath model. 

The current profile obtained using the multi-sheath model [Eq.~(\ref{eq:lambda_exact})] is shown in Fig.~\ref{fig:beam_loading_tz}(b) in solid red. For reference, we also plot the beam profile $\lambda(\xi) \simeq 6.96 - 1.75 \times (\xi - 8.27)$ [Eq.~(\ref{eq:lambdaopt})] in the ultrarelativistic limit in dashed blue. While both models predict trapezoidal profiles with similar slice currents, the bunch length predicted by the multi-sheath model $\Delta \xi_{tr} \simeq 3.09$ is nearly $\sim 42 \%$ longer than that of the analytic theory $\Delta \xi_{tr} = \frac{r_t^2}{4E_t} \simeq 2.18$. Despite the fact that most of the charge is front-loaded in both profiles, the total charge predicted by the multi-sheath model $Q_{tr}/en_pk_p^{-3} \simeq 87.1$ is still $\sim 25 \%$ more than the total charge predicted by the analytic theory $Q_{tr}/en_pk_p^{-3} \simeq 69.4 $. 

The beam profile obtained using the single-sheath model (dashed green) with $n_2 = 0$ is also shown for qualitative comparisons. Since the head of the bunch is situated at $r_t/r_m \simeq 0.78$ where the second sheath can be largely neglected, the single-sheath model will initially predict slice currents similar to those obtained using the multi-sheath model. However, at lower $\tilde{r}_b$, the multi-sheath model and single-sheath begin to diverge as the second sheath comes into play. Eventually, the current profile predicted by the single-sheath model turns negative $(\lambda < 0)$, similar to what can be seen in Fig.~\ref{fig:beam_loading}(b), due to the absence of the characteristic spike in the electric field when using the single-sheath model.

 In Fig.~\ref{fig:beam_loading_tz}(d), we plot the PIC simulation results using the profiles given by the multi-sheath model and analytic theory shown in Fig.~\ref{fig:beam_loading_tz}(c). It can be readily seen that the multi-sheath model provides improved accuracy over the analytic theory in terms of flattening the wakefield. Furthermore, the beam load predicted by the multi-sheath extends all the way to the very rear of the wake while the beam load predicted by the analytic theory does not.

\subsection{Total accelerating force}

When loading a constant wakefield $f(\xi \geq \xi_t) = -E_t$, the interplay between the maximum loaded charge $Q_{tr}$ and the accelerating field $E_t$ can be characterized by examining the total accelerating force $Q_{tr}E_t$.  In Ref.~\citenum{tzoufrasprab}, the total accelerating force was found to be
\begin{align}
\label{eq:tzqtet}
\frac{Q_{tr}}{e} \frac{eE_t}{mc^2/r_e} = \frac{1}{4^3} (k_p r_m)^4,
\end{align}
in the ultrarelativistic limit by integrating the analytic theory described by Eq.~(\ref{eq:lambdaopt}). An exact calculation for $Q_{tr}E_t$ can be obtained by numerically integrating loaded trajectory $\tilde{r}_b(\xi)$ from Eq.~(\ref{eq:efieldloaded}) and, then, integrating the current profile $\lambda(\xi)$ described by Eq.~(\ref{eq:lambda_exact}) using
\begin{align}
\label{eq:qtet_exact}
\frac{Q_{tr}}{e} \frac{eE_t}{mc^2/r_e} = \frac{1}{2} \int^{\xi_f}_{\xi_t} \lambda(\xi)E_td\xi 
\end{align}
for a specified constant wakefield $E_t$ until the very rear of the wake defined by $\tilde{r}_b(\xi_f) = 0$.

\begin{figure}[t]
\includegraphics[width=0.5\textwidth]{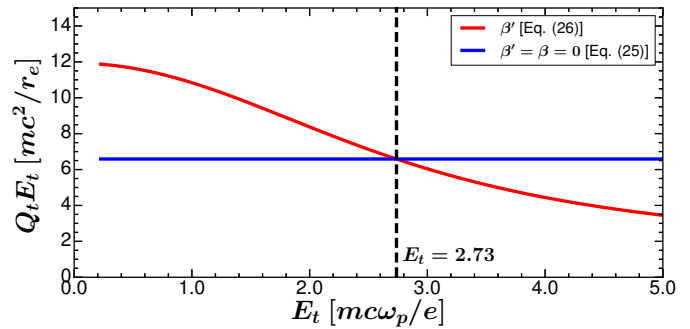}
\caption{\label{fig:qtet} The accelerating force $Q_t E_t$ as a function of the accelerating field $E_t$ for the plasma wakefield from Fig.~\ref{fig:beam_loading} with maximum bubble radius $r_m \simeq 4.53$.  The blue curve corresponds to Eq.~(\ref{eq:tzqtet}). The red curve is numerically integrated from Eq.~(\ref{eq:qtet_exact}) using the multi-sheath model $\beta^{\prime}$ [Eq.~(\ref{eq:lambda_exact})] to load constant wakefields, i.e., $E_z(\xi \geq \xi_t) =  -E_t$, at different positions in the wake. The same integration parameters $\{\Delta_{10}=0.875, \Delta_{20} = 3, s = 3, \psi_{min} = -1\}$ from Fig.~\ref{fig:beam_loading} are used for each calculation. }
\end{figure}

In Fig.~\ref{fig:qtet}, we plot the accelerating force $Q_{tr}E_t$ as a function of the accelerating field $E_t$ for the plasma wakefield shown in Fig.~\ref{fig:beam_loading}(a) with a blowout radius $r_m \simeq 4.53$. The blue line corresponds to the analytic theory described by Eq.~(\ref{eq:tzqtet}) while the red line is obtained by numerically integrating Eq.~(\ref{eq:qtet_exact}) using the multi-sheath model starting at different positions in the wake. For simplicity, we use $\psi_{min} = -1$ to calculate $Q_{tr}E_t$ using the multi-sheath model rather than tailoring $\psi_{min}$ for cases with long beam loads, i.e., $\Delta \xi_{tr} \gtrsim r_m/2$. As shown in the previous section, adjusting $\psi_{min}$ to account for self-consistent beam loading effects can decrease the predicted charge by $\lesssim O(10\%)$.

As can be seen in Fig.~\ref{fig:qtet}, Eq.~(\ref{eq:tzqtet}) predicts a constant accelerating force regardless of where the load is placed while the multi-sheath model predicts an accelerating force that decreases as the amplitude of the accelerating field $E_t$ increases. For values of $E_t < 2.73$, the multi-sheath model predicts more loaded charge $Q_{tr}$ than the analytic theory due to its longer beam loads with comparable slice currents. It is worth pointing out that all examples of beam loading presented in Figs.~\ref{fig:beam_loading} and \ref{fig:beam_loading_tz} were operating in this range.  On the other hand, the multi-sheath model predicts less charge can be loaded for larger accelerating fields $E_t > 2.73$ because the lower slice currents now outweigh the differences between the predicted bunch lengths. While the exact crossing point $E_t$ will vary on a case by case basis, the qualitative features of the accelerating force $Q_t E_t$ predicted by multi-sheath model will be largely similar for nonlinear wakes with different $r_m$. Since the analytic theory is also a limit of the multi-sheath model, the accelerating force $Q_{tr}E_t$ obtained using the multi-sheath model also scales with $r_m^4$ when $r_m \gg 1$.

\subsection{Beam loading longitudinally varying wakefields}

\begin{figure}[t]
\includegraphics[width=0.5\textwidth]{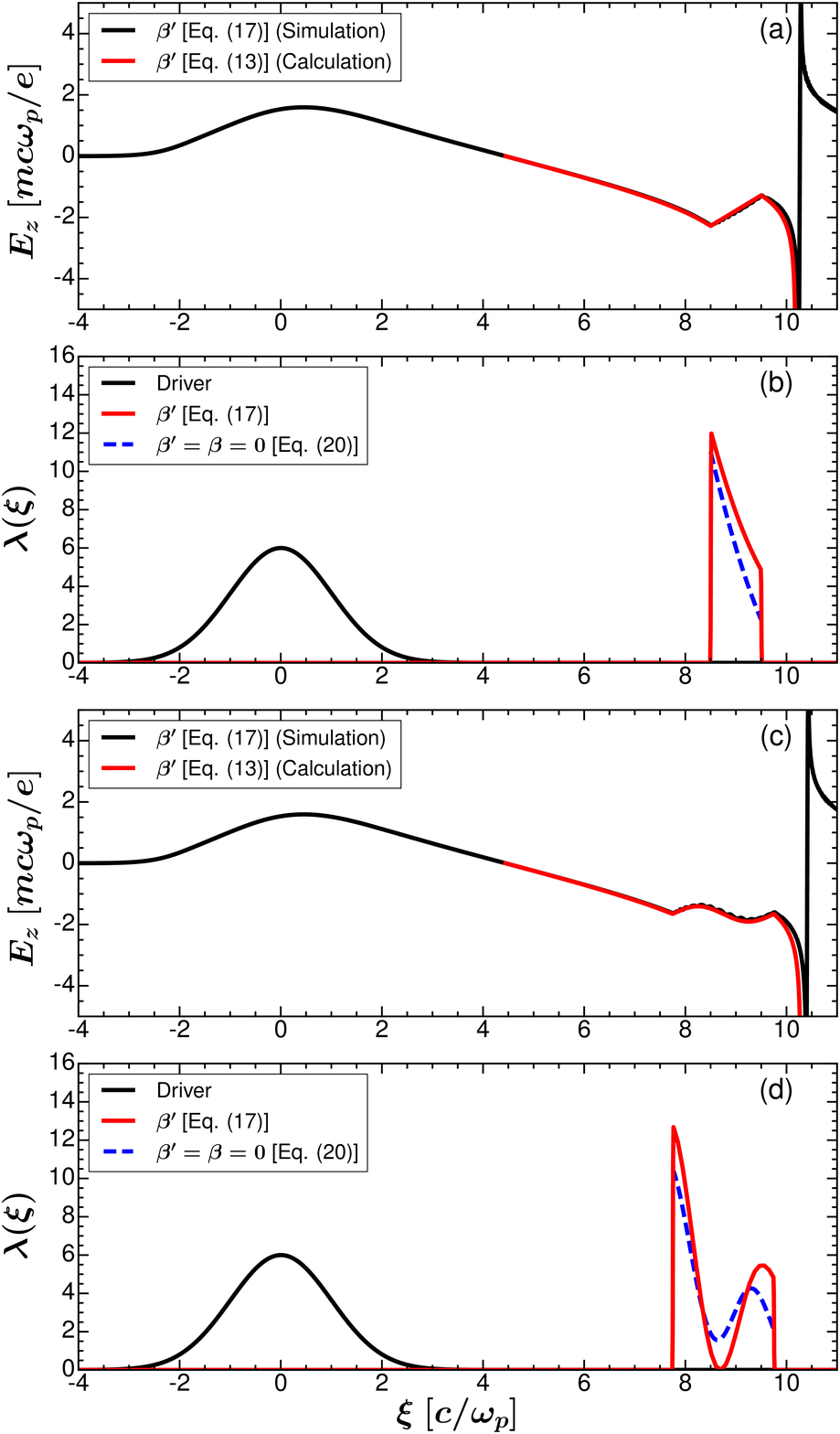}
\caption{\label{fig:arb_functions} Simulation results (black) using the multi-sheath model [Eqs.~(\ref{eq:efieldloaded})-(\ref{eq:lambda_exact})] with $\psi_{min} = -1$ for two different electric field profiles: (a) $f(\xi_t \leq \xi \leq \xi_f) = -E_t + (\xi - \xi_t)$ from $\xi_t = 8.5$ to $\xi_f = 9.5$ and (c) $f(\xi_t \leq \xi \leq \xi_f) = -E_t + E_{1} \sin(k(\xi - \xi_t))$ from $\xi_t = 7.75$ to $\xi_f = 9.75$ where $E_{1} = \frac{1}{4}$ and $k = \pi$. The dashed red lines correspond to the electric field calculated from Eq.~(\ref{eq:efield}). The corresponding current profiles $\lambda(\xi)$ [Eq.~(\ref{eq:lambda_exact})] from (a) and (c) are plotted in (b) and (d), respectively. Analytic current profiles $\lambda(\xi)$ calculated from Eq.~(\ref{eq:lambda_simplified}) are plotted in dashed blue.  }
\end{figure}

While much of the discussion has focused on loading constant wakefields, we will now present several examples in which longituindally varying plasma wakefields are loaded using the formalism described by Eqs.~(\ref{eq:efieldloaded})-(\ref{eq:lambda_exact}) of the multi-sheath model. Designing beam loads for tailored $\frac{dE_z}{d\xi}$ is of interest to self-injection or collider designs where initial energy chirps are present in the witness beam. Beam profiles will also be presented in the ultrarelativistic limit $(\beta^{\prime},\beta \rightarrow 0,0)$ using Eqs.~(\ref{eq:efieldloadedsimplified}) and (\ref{eq:lambda_simplified}). In each case, an electron drive bunch $\{ \Lambda_d=6, \gamma_b = 20000, k_p \sigma_r = 0.245, k_p \sigma_z = 1, k_p \xi_c = 0 \}$ is used to excite a plasma wake with a blowout radius $r_m \simeq 4.53$. Since the driver is identical to the one used in Fig.~\ref{fig:beam_loading}, we will also use the same parameters $\{\Delta_{10} = 0.875, \Delta_s = 0.05r_b, \Delta_{20} = 3, s = 3\}$ to obtain results using the multi-sheath model $\beta^{\prime}$. $n_{20}$ is calculated from Eq.~(\ref{eq:n2_limit}) using $\psi_{min} = -1$ since the trailing bunches are short. For this driver, we refer to Fig.~\ref{fig:beam_loading}(a) for the unloaded wake trajectory (dashed red) calculated using the multi-sheath model.

In Figs.~\ref{fig:arb_functions}(a)-(b), we design a beam profile that loads a linear plasma wakefield with a positive slope $df/d\xi > 0$. Such a wakefield can be used to dechirp a beam with a positive energy chirp $d_{\xi} \gamma > 0$ while still maintaining an accelerating field $E_z <0$ over the electron load. In this example, we choose to load the function $f(\xi) = -E_t +(\xi - \xi_t)$ from $\xi_t = 8.5$ to $\xi_f = 9.5$ such that $df/d\xi = 1$. From the unloaded trajectory calculated from Eq.~(\ref{eq:rbeq}), the bubble radius $r_t \simeq 2.73$ and electric field $E_t \simeq 2.26$ at the head of the bunch are known. 

To calculate the beam load profile, we must first numerically integrate Eq.~(\ref{eq:efieldloaded}) to obtain the loaded bubble trajectory $\tilde{r}_b(\xi)$ from $\xi_t$ to $\xi_f$. After the beam load $\xi > \xi_f$, the remaining unloaded trajectory is calculated by integrating Eq.~(\ref{eq:rbeq}) starting at $\tilde{r}_b(\xi_f)$. The electric field is then calculated from Eq.~(\ref{eq:efield}) across all regions (unloaded and loaded) and plotted (solid red) in Fig.~\ref{fig:arb_functions}(a). From the results, it is clearly evident that the loaded region does not extend to the very rear of the wake since the characteristic negative spike in the electric field is still present. In Fig.~\ref{fig:arb_functions}(b), the underlying current profile $\lambda(\xi)$ calculated from Eq.~(\ref{eq:lambda_exact}) is shown (solid red). Using this current profile, PIC simulation results (solid black) shown in Fig.~\ref{fig:arb_functions}(a) indeed confirm that the desired wakefield $f(\xi)$ is produced along the bunch. Strong agreement is also observed between the simulated and calculated electric fields in regions before and after the beam load. 

In the ultrarelativistic limit $\beta^{\prime} \rightarrow 0$, it is straightforward to show that the analytic solution to Eq.~(\ref{eq:efieldloadedsimplified}) along the beam load is a hyperbola $\tilde{r}_b(\xi)^2 = r_t^2- 4E_t(\xi -\xi_t) + 2(\xi -\xi_t)^2$. Substituting $\tilde{r}_b$ into Eq.~(\ref{eq:lambda_simplified}), the analytic current profile is a parabola $\lambda(\xi) = \frac{3}{4}r_t^2+ E_t^2- 5E_t(\xi-\xi_t) + \frac{5}{2}(\xi-\xi_t)^2$. This profile is evaluated using $r_t \simeq 2.81$ and $E_t \simeq 2.26$ from the PIC simulation data and plotted (dashed blue) in Fig.~\ref{fig:arb_functions}(b). While this analytic profile captures the general trend of the multi-sheath results, we note that disagreement is still observed between the two profiles along portions of the load. 

In Figs.~\ref{fig:arb_functions}(c)-(d), we design a beam profile that loads a sinusoidally oscillating plasma wakefield to highlight the limitations of the loaded plasma wakefield slope $df/d\xi$. For this case, we choose to load the function $f(\xi) = -E_t + E_1 \sin(k(\xi-\xi_t))$ from $\xi_t = 7.75$ to $\xi_f = 9.75$ where $E_1 = \frac{1}{4}$ and $k = \pi$ . From the unloaded bubble trajectory calculated from Eq.~(\ref{eq:rbeq}), we use $r_t \simeq 3.45$ and $E_t \simeq 1.66$ to numerically integrate the loaded bubble trajectory $\tilde{r}_b(\xi)$ [Eq.~(\ref{eq:efieldloaded})] from $\xi_t$ to $\xi_f$. After the beam load $\xi > \xi_f$, the remaining unloaded trajectory is calculated from Eq.~(\ref{eq:rbeq}) starting at $\tilde{r}_b(\xi_f)$. In Fig.~\ref{fig:arb_functions}(c), we plot the electric field calculated from Eq.~(\ref{eq:efield}) using the multi-sheath model in solid red. The current profile of the load $\lambda(\xi)$ calculated from Eq.~(\ref{eq:lambda_exact}) is plotted (solid red) in Fig.~\ref{fig:arb_functions}(d). Using this current profile, the electric field from PIC simulation results (solid black) exhibits strong agreement with the multi-sheath results in all regions. 

As pointed out previously, $\lambda(\xi)$ is defined to be positive definite for an electron load, which limits how negative $df/d\xi$ can be in Eq.~(\ref{eq:lambda_exact}). As shown in Figs.~\ref{fig:arb_functions}(c)-(d), the current profile approaches zero around $\xi \approx 8.7$ where $df/d\xi$ is near its minimum. In this case, increasing the amplitude or the frequency of the sinusoidal oscillation would result in a more negative slope $df/d\xi$, which would require a positive (positron) charge density along regions of the load to attract the sheath electrons that trace the bubble trajectory back to the axis more quickly. 

A similar analysis can be done in the ultrarelavistic limit where the analytic solution to Eq.~(\ref{eq:efieldloadedsimplified}) is $\tilde{r}_b(\xi)^2 = r_t^2 -4 E_t(\xi- \xi_t) +\frac{4E_1}{k} \big[ 1- \cos(k(\xi - \xi_t)) \big]$. Using this analytic trajectory, the current profile of the underlying bunch $\lambda(\xi) = \frac{\tilde{r}_b^2}{4}+ f(\xi)^2 +\frac{\tilde{r}_b^2}{2}\frac{df}{d\xi}$ [Eq.~(\ref{eq:lambda_simplified})] can now be completely expressed in terms of $\xi$ where $\frac{df}{d\xi} = kE_1\cos(k(\xi- \xi_t))$. The slope $df/d\xi$ is naturally constrained since it is the only term which can be negative and $\lambda \geq 0$ for an electron driver by definition. In Fig.~\ref{fig:arb_functions}(c), this profile is evaluated using $r_t \simeq 3.49$ and $E_t \simeq -1.63$ from unloaded PIC simulation results and plotted (dashed blue). While the analytic current profile qualitatively reproduces the oscillations observed in the multi-sheath profile, it is still an approximation of the multi-sheath model and, therefore, deviates from it along portions of the beam load. For example, near $\xi \approx 8.7$, the analytic current profile dips to $\lambda \approx 1.56$ whereas the multi-sheath profile approaches $\lambda \approx 0$.  In Sec.~\ref{sec:discussion}, we provide more detailed comparisons between the single and multi-sheath models and the analytic results. Explanations for these differences are also given.

\section{Beam loading in laser wakefields}

Up to this point, we assumed that the wakefields are excited by electron drivers. However, the formalism described in Sec.~\ref{sec:potential} can be easily extended to a laser driver specified by the vector potential $A_{laser} = \Re\{A_{\perp} e^{i\omega_0/c\xi} \}$ where $\omega_0$ is the laser frequency and $a \equiv  eA_{\perp}/mc^2$ is the normalized vector potential envelope. To do this, we use the same source term profile for $S$ described by Eq.~(\ref{eq:sprofile}). Therefore, the expressions for the wake potential $\psi= (1+\beta^{\prime})r_b^2/4 - r^2/4$ obtained by integrating Eq.~(\ref{eq:delpsi}) and the electric field $E_z = \frac{d\psi_0}{d\xi} = D^{\prime}(r_b) r_b \frac{dr_b}{d\xi}$ are identical to the those derived in Sec.~\ref{sec:potential}. The main difference for a laser driver is that the plasma electrons are now displaced by the ponderomotive force

\begin{align}
\label{eq:ponderomotive}
\bold{F}_p = -\frac{1}{\bar{\gamma}} \nabla \frac{|a|^2}{4}
\end{align}
where $\bar{\gamma} = \sqrt{1 + P^2 + |a|^2/2}$ \cite{mora}. As shown in Ref.~\citenum{lu2006nonlinearphysplasma}, the total transverse force on the sheath electron that traces $r=r_b(\xi)$ can now be written as
\begin{align}
\label{eq:totalforce}
F_{\perp} = &-\frac{1}{2}r + (1-v_z)\left[-\frac{1}{2}\frac{d^2\psi_0}{d\xi^2}r\right]  \notag \\
& + (1-v_z) \frac{\lambda(\xi)}{r} -\frac{1}{\bar{\gamma}} \nabla \frac{|a|^2}{4} 
\end{align}
where the first term is the linear focusing force due to the ions, the second term is the force from the radial sheath currents of the plasma, the third term is the defocusing force due to a trailing bunch with a current profile $\lambda(\xi)$, and the fourth term is the ponderomotive force of the laser driver.  From the constant of motion $\bar{\gamma} - P_z = 1 + \psi$ \cite{mora}, it can also be shown that $1-v_z = \frac{2(1+\psi)^2}{1+ P_{\perp}^2 + |a|^2/2 + (1+\psi)^2}$. Furthermore, the relativistic equation of motion of the plasma electron that traces $r_b(\xi)$ can also be expressed as

\begin{align}
\label{eq:momentum}
\frac{dP_{\perp}}{d\xi} = \frac{d}{d\xi} \left[(1+\psi) \frac{dr_b}{d\xi} \right] = \frac{1}{1-v_z} F_{\perp}. 
\end{align}

Substituting Eq.~(\ref{eq:totalforce}) into the right-hand side of Eq.~(\ref{eq:momentum}), we obtain
\begin{align}
\label{eq:rbgenerallaser}
\frac{d}{d\xi} \Bigg[ (1+\psi) &\frac{d}{d\xi}r_b \Bigg] = r_b \Bigg\{ -\frac{1}{4} \Bigg[1 + \frac{1+|a|^2/2}{(1+\psi)^2} + \left(\frac{dr_b}{d\xi}\right)^2 \Bigg] \notag \\
&- \frac{1}{2} \frac{d^2\psi_0}{d\xi^2}  + \frac{\lambda(\xi)}{r_b^2 }   \Bigg\} -\frac{1}{(1+\psi)} \nabla_{\perp} \frac{|a|^2}{4}.
\end{align}
Assuming $\beta^{\prime}$ is an explicit function of $r_b(\xi)$, i.e., $d\beta^{\prime}/d\xi = \partial \beta^{\prime}/\partial r_b (dr_b/d\xi)$, the trajectory of the sheath electron for a laser driver, as derived by Lu et al.~\cite{lu2006nonlinearphysplasma}, can be rewritten as
\begin{align}
\label{eq:rbeqlaser}
A^{\prime}(r_b) \frac{d^2r_b}{d\xi^2} &+ B^{\prime}(r_b) r_b \left( \frac{dr_b}{d\xi}\right)^2 +C_L^{\prime}(r_b)r_b\notag \\
& = \frac{\lambda (\xi)}{r_b}-G_L^{\prime}(r_b) \nabla_{\perp} |a|^2 
\end{align}
where the new coefficients for the laser case denoted with the subscript ``L"  are defined as
\begin{align*}
C_L^{\prime}(r_b)  &= \frac{1}{4} \left[\vcenter{\hbox{$\displaystyle 1+ \cfrac{1+|a|^2/2}{\left(1+\cfrac{\beta^{\prime}r_b^2}{4}\right)^2}      $}}\right], \notag \\
\notag \\
G_L^{\prime}(r_b)  &= \frac{1}{4} \left[\vcenter{\hbox{$\displaystyle \cfrac{1}{\left(1+\cfrac{\beta^{\prime}r_b^2}{4}\right)}      $}}\right]. \notag \\
\end{align*}

\begin{figure*}[t]
\includegraphics[width=1\textwidth]{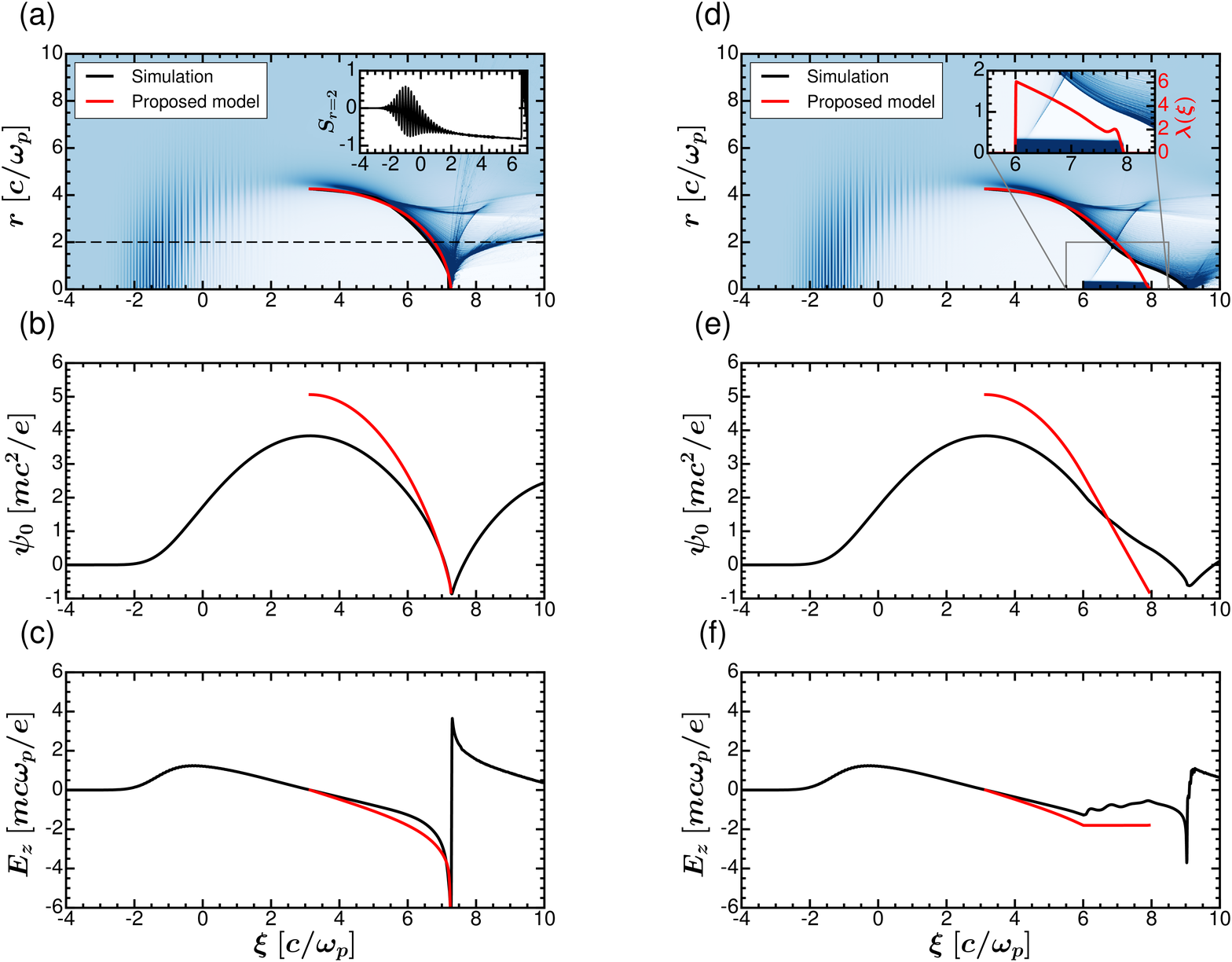}
\caption{\label{fig:lasercase} Comparisons of simulations (black) and numerical calculations (red) of $r_b$, $\psi_0$ and $E_z$ using Eqs.~(\ref{eq:psi_final}), (\ref{eq:efield}), and (\ref{eq:rbeqlaser}) for laser wakefields excited by a 40 fs (FWHM) 0.8 $\mu \text{m}$ laser driver after propagating a distance $z= 0.32~ \text{mm}$ into a plasma with density $n_p = 1.5 \times 10^{18} ~\text{cm}^{-3}$. In both cases, the laser pulse is initially focused at the plasma entrance with a normalized vector potential $a_0 = 4$, normalized spot size $k_pW_0 = 4$, and centroid $k_p\xi_c = 0$. At $z=0.32~ \text{mm}$, the electron density distribution is shown for two cases with (a) no load and (d) a beam load designed to produce a constant wakefield starting at $k_p\xi = 6$ using Eqs.~(\ref{eq:efieldloaded}) and (\ref{eq:lambda_exact_laser}). The insets show the lineout of the source term $S$ at $k_pr= 2$ (dashed black) for the unloaded case and the trailing bunch profile $\lambda(\xi)$ (red) obtained using the multi-sheath model for the loaded case. The integration parameters are $\Delta_{10}=0.3$, $\Delta_{20}=3$ , and $s=3$. $n_{20}$ is calculated from Eq.~(\ref{eq:n2_limit}) using $\psi_{min}= -0.85$ and $n_1$ is calculated from Eq.~(\ref{eq:constant1_eval2}).  }
\end{figure*}

It is worth mentioning that $A^{\prime}(r_b)$ and $B^{\prime}(r_b)$ are the same coefficients specified for the beam-driven wake in Sec.~\ref{sec:potential}.  While the equation of motion describing the trajectory of the sheath electron is slightly different in the case of a laser driver, the general procedure for modeling the wakefield remains the same. For the laser wakefields in this section, we use the same profiles for $\Delta_{1}(r_b)$, $\Delta_{2}(r_b)$ and $n_2(r_b)$ as described in Sec.~\ref{sec:potential}. As in the beam-driven case, $n_1(r_b)$ is calculated from Eq.~(\ref{eq:constant1_eval2}) and $n_{20}$ is constrained by Eq.~(\ref{eq:n2_limit}). Using these quantities, we can calculate $\beta^{\prime}(r_b)$ from Eq.~(\ref{eq:beta}) and, then, numerically integrate Eq.~(\ref{eq:rbeqlaser}) to obtain the trajectory of the sheath electron that traces $r_b(\xi)$ starting at the maximum bubble radius $r_m$. Once $r_b(\xi)$ is known, the wake potential $\psi(r,\xi)$ and electric field $E_z(\xi)$ can be obtained everywhere inside the bubble $r < r_b(\xi)$.  

Determining the exact value of the blowout radius $r_m$ for a laser wakefield is generally more difficult because the electrons are not completely blown out by the ponderomotive force of the laser, which is largely localized to the laser spot size $W_0$. As a result, the particle tracing ``$r_b$" is no longer the innermost electron but the characteristic sheath electron with the largest forward velocity $v_z = 1- \frac{2(1+\psi)^2}{1+ P_{\perp}^2 + |a|^2/2 + (1+\psi)^2}$ near the axis where $\psi$ is minimum. Like in the beam-driven cases, the minimum wake potential $\psi_{min}$ can be well-approximated by values close to $-1$ when $r_m \gtrsim 3$ for the trajectory traced out by this electron. This regime typically corresponds to lasers with normalized vector potentials $a_0 \gtrsim 2$ since $r_m \approx 2\sqrt{a_0}$ (which is only valid if $a_0 \gtrsim 2$).

The methodology for loading a wakefield $E_z(\xi_t \leq \xi \leq \xi_f) = f(\xi)$ also remains largely unchanged from the procedure described in Sec.~\ref{sec:beamloading}. By integrating Eq.~(\ref{eq:efieldloaded}), we can then obtain the modified sheath electron trajectory $\tilde{r}_b(\xi)$ corresponding to a loaded wakefield $f(\xi)$ starting at the beam head located at $\xi_t$. The only difference is that the corresponding current profile for the beam load is now given by 

\begin{align}
\label{eq:lambda_exact_laser}
\lambda(\xi) =&  C_L^{\prime}\tilde{r}_b^2 + \left(\frac{B^{\prime}}{D^{\prime2}} -\frac{A^{\prime}F^{\prime} }{D^{\prime3} \tilde{r}_b^2} \right)f(\xi)^2  \notag \\
&+\left(\frac{A^{\prime}}{D^{\prime}}\right)\frac{df(\xi)}{d\xi} + G_L^{\prime}\tilde{r}_b \nabla_{\perp} |a|^2 
\end{align}
where the last term corresponds to the ponderomotive force from Eq.~(\ref{eq:rbeqlaser}). In cases with short laser pulses, the ponderomotive term in Eq.~(\ref{eq:rbeqlaser}) can be dropped in the back half of the wake.

In Figs.~\ref{fig:lasercase}(a)-(c), we show the results for the bubble trajectory $r_b(\xi)$, potential $\psi_0(\xi)$, and electric field $E_z(\xi)$ obtained from calculations using the multi-sheath model (red) and {\scshape osiris} (quasi-3D) PIC simulation results (black) for an unloaded wake excited by a 40 fs (FWHM) 0.8 $\mu$m laser driver after propagating into a constant plasma density $n_p = 1.5 \times 10^{18} ~\text{cm}^{-3}$. The PIC simulation used a customized finite-difference solver to reduce numerical effects from relativistic particles \cite{xu2020,li2017}, a high resolution grid $\Delta r = \Delta z= \frac{1}{128} \frac{c}{\omega_p}$ with $\Delta t = \frac{1}{512} \frac{1}{\omega_p}$, and 32 particles per cell (2x2x8). The laser is initially focused at the plasma entrance with a normalized vector potential $a_0 = 4$ and a transverse gaussian envelope having a matched spot size $k_p W_0 = 2\sqrt{a_0} = 4$ \cite{lu2006nonlinearphysplasma}. The electron density distribution in the $r-\xi$ plane is shown in Fig.~\ref{fig:lasercase}(a) after a propagation distance $z=$ 0.32 mm into the plasma at which point the blowout radius is $r_m \simeq 4.26$. It can readily be seen that while the multi-sheath model generates a sheath electron trajectory $r_b(\xi)$ in Fig.~\ref{fig:lasercase}(a) that is in good agreement with the bubble trajectory from the simulation results, it overestimates (underestimates) the potential (electric field) over most of the wake in Figs.~\ref{fig:lasercase}(b)-(c). The disagreement is primarily due to the fact that the plasma electrons are not completely blown out by the ponderomotive force of the laser. Therefore, plasma electrons can now propagate inside the bubble, i.e., $r<r_b$, resulting in spatially varying charge densities and currents in the bubble.

As a result, the expressions for $\psi(r,\xi)$ [Eq.~(\ref{eq:psi_final})] and $E_z(\xi)$ [Eq.~(\ref{eq:efield})] obtained using the multi-sheath model [Eq.~(\ref{eq:sprofile})] break down because the source term inside the channel $S(r<r_b)$ is no longer exactly $-1$ as shown in the inset plot of $S$ along $r=2$ in Fig.~\ref{fig:lasercase}(a). The presence of plasma electrons inside the bubble is also important because the focusing force is no longer perfectly linear. In addition, these electrons can move from the inside $(r<r_b)$ to the outside of the bubble $(r > r_b)$ effectively splitting the plasma sheath into two. This effect can be seen from the simulation results in the Fig.~\ref{fig:lasercase}(a) inset near the bubble radius $r_b(\xi)$ at $\xi \approx 6$. This sheath splitting phenomenon can typically produce more than 3 distinct regions in which $S$ has alternating signs, which differs from the model assumed in Eq.~(\ref{eq:sprofile}).

In Figs.~\ref{fig:lasercase}(d)-(f), we show the electron density distribution of the laser-driven wakefield at $z=$ 0.32 mm with a beam load designed to produce a constant wakefield starting at $\xi_t = 6$. To self-consistently load the wake using Eqs.~(\ref{eq:efieldloaded}) and (\ref{eq:lambda_exact_laser}), the electric field at the head of the bunch is sampled from the multi-sheath model $f(\xi > \xi_t) = -E_t = -1.74$ rather than the simulation results. The current profile $\lambda(\xi)$ of the beam load calculated from Eq.~(\ref{eq:lambda_exact_laser}) is shown in the inset of Fig.~\ref{fig:lasercase}(d). 

It is clear from the simulation results that the multi-sheath model fails to capture the behavior of the modified sheath electron trajectory $\tilde{r}_b$ [Eq.~(\ref{eq:efieldloaded})], wake potential $\psi_0$, and electric field $E_z$ in Figs.~\ref{fig:lasercase}(d)-(f). The underlying reason is that the multi-sheath model predicts an electric field at the head of the bunch that is more negative than the simulated electric field. Therefore, the currents calculated from Eq.~(\ref{eq:lambda_exact_laser}) are larger than needed due to the $f(\xi)^2$ term on the right-hand side. The simulation results show that the current profile produces an electric field that actually increases along the beam load rather than remaining constant. 

It is also worth noting that the beam load blows out the remaining electrons inside the channel $r < r_b$ and forms another thin plasma sheath as can be seen in the electron density phase space and inset plots in Fig.~\ref{fig:lasercase}(d). As these electrons are being blown out, the source term inside the bubble becomes more negative until only ions remain and $S(r < r_b) = -1$. This effect also contributes to the positive slope of the loaded wakefield $E_z$ near the head of the bunch $\xi_t$.

From the results presented in this section, it is evident that the model for $S$ described by Eq.~(\ref{eq:sprofile}) is not sufficient for modeling unloaded and loaded laser wakefields. Plasma electrons propagating inside the ``bubble," sheath splitting, and blowout of remaining electrons by the beam load are some of the features making it difficult to apply the multi-sheath model, as is, to cases with a laser driver. For these very same reasons, electron beams are ideal for driving high-quality plasma wakefields in which electrons are completely blown out and the focusing force is perfectly linear. While the multi-sheath model can be adapted to laser drivers by using a source term model in which $S(r< r_b)$ is no longer constant, the force in Eq.~(\ref{eq:totalforce}) will also need to be modified due to the fields from the plasma currents inside the bubble. Such an analysis will also require assumptions about the electron currents inside the channel. This is an area for future work.

\section{Differences between the sheath models and analytic theory for beam loading}
\label{sec:discussion}

In this section, we provide details regarding the differences in the predictions between the sheath and analytical models. These details also show why the anlaytic model provides reasonable agreement for the witness beam current but poor predictions for $r_b$. 

From the results presented in Sec.~\ref{sec:beamloading}, it is clear that the analytic theory can be a useful tool for predicting the general form of the current profiles for beam loading. However, as it is an approximation of the multi-sheath (and single-sheath) model, it is generally not as accurate even for $r_b \sim r_m$. For beam loads designed to produce constant wakefields, the resulting parabolic trajectory $\tilde{r}_b(\xi)^2 = r_t^2 - 4E_t(\xi - \xi_t)$ predicted by Eq.~(\ref{eq:efieldloadedsimplified}) can also deviate significantly from that of the multi-sheath model as seen in Figs.~(\ref{fig:beam_loading})-(\ref{fig:beam_loading_tz}). As a result, the analytic theory can underestimate the maximum length of the beam load $\Delta \xi_{tr}$ and, thus, the total charge $Q_{tr}$ when compared to the multi-sheath results for these cases. Despite this, the slice currents predicted by the analytic theory are still comparable to those obtained using the multi-sheath model. 

To understand why this occurs, we revisit the differential equation for the bubble trajectory from Eq.~(\ref{eq:rbeq}). In the ultrarelativistic limit $(\beta^{\prime},\beta \rightarrow 0,0)$, Eq.~(\ref{eq:rbeq}) describing the innermost particle trajectory $r_b(\xi)$ was found to be \cite{tzoufrasprab,tzoufrasprl}

\begin{align}
\label{eq:rbtz}
r_b \frac{d^2r_b}{d\xi^2} + 2 \left( \frac{dr_b}{d\xi}\right)^2 + 1 = \frac{4\lambda (\xi)}{r_b^2}
\end{align}
where the wake potential is now 
\begin{align}
\label{eq:psitz}
\psi_0(\xi) \approx \Psi_{\text{I}} = \frac{r_b^2}{4}
\end{align}
and the electric field is $E_z(\xi) = \frac{d\psi_0}{d\xi} = \frac{1}{2} r_b \frac{dr_b}{d\xi}$. 
As shown in Refs.~\citenum{tzoufrasprl} and \citenum{tzoufrasprab}, Eq.~(\ref{eq:rbtz}) can be integrated starting at the blowout radius $r_m$ to obtain the following expression for the bubble trajectory for an unloaded plasma wake $(\lambda = 0)$

\begin{figure}[t]
\includegraphics[width=0.5\textwidth]{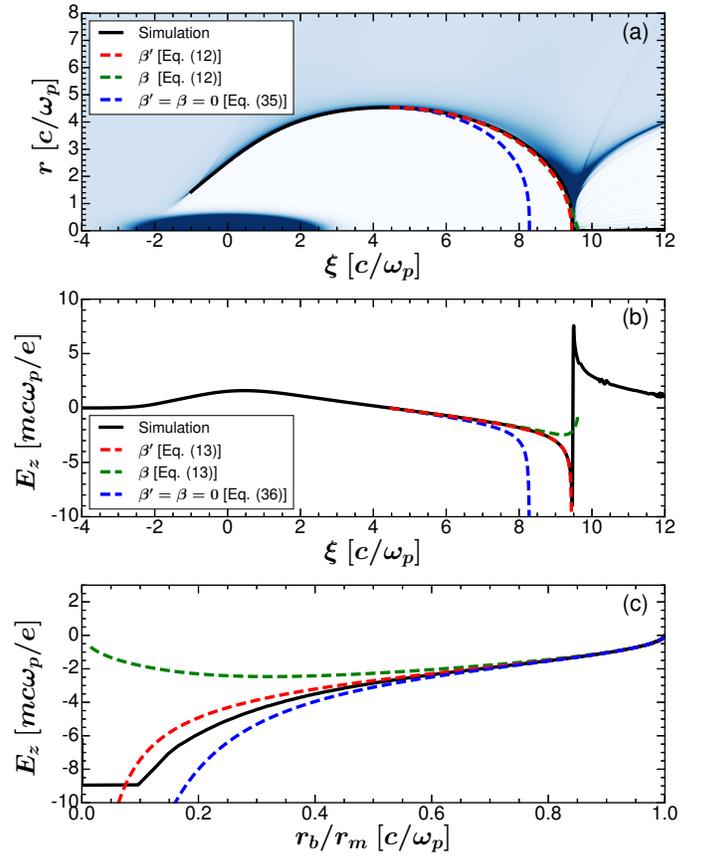}
\caption{\label{fig:tzoufras_diff} (a) Electron density distribution and electric field $E_z$ as a function of (b) $\xi$ and (c) $r_b(\xi)$ for a plasma wake excited by a nonevolving driver with parameters $\Lambda_d = 6$, $\gamma_b = 20000$, $k_p \xi_c = 0$, $k_p\sigma_r =0.245$, and $k_p\sigma_z =1$. The maximum bubble radius is $k_pr_m \simeq 4.53$. Simulation data is shown in black. Numerically integrated results are shown for the multi-sheath model $\beta^{\prime}$ (dashed red) with $\psi_{min} = -1$ and the single-sheath model $\beta$ (dashed green) with $\psi_{min} = 0$ using Eqs.~(\ref{eq:psi_final})-(\ref{eq:n2_limit}) with $\Delta_{10} = 0.875$, $\Delta_s = 0.05r_b$, $\Delta_{20} = 3$, and $s =3$. Analytic results using Eqs.~(\ref{eq:xirbtz}) and (\ref{eq:efieldtzunloaded}) are shown in dashed blue. }
\end{figure}

\begin{align}
\label{eq:xirbtz}
\frac{\xi}{r_m} = 2E\left[ \arccos \left( \frac{r_b}{r_m}  \right)\bigg| \frac{1}{2} \right] -F\left[ \arccos \left( \frac{r_b}{r_m}  \right)\bigg| \frac{1}{2} \right] \notag \\
\end{align}
where $F$ and $E$ are incomplete elliptical integrals of the first and second kind. The corresponding electric field can also be calculated analytically to be \cite{tzoufrasprab,tzoufrasprl}

\begin{align}
\label{eq:efieldtzunloaded}
E_z(\xi) = \frac{1}{2} r_b \frac{dr_b}{d\xi} = -\frac{r_b}{2\sqrt{2}} \sqrt{\frac{r^4_m}{r_b^4} -1}.
\end{align}

In Fig.~\ref{fig:tzoufras_diff}, we compare the analytic theory [Eqs.~(\ref{eq:xirbtz})-(\ref{eq:efieldtzunloaded})] with the multi-sheath model ($\beta^{\prime}$), single-sheath model ($\beta$), and simulation results for an unloaded plasma wake excited by an electron driver with $\Lambda_d = 6$, $k_p\sigma_z = 1$, $\gamma_b = 20000$, and $k_p\sigma_r = 0.245$. For the multi-sheath and single-sheath calculations, the integration parameters are specified in the figure caption. From Fig.~\ref{fig:tzoufras_diff}(a), it is clear that the bubble trajectory $r_b(\xi)$ (blue dashed) described by Eq.~(\ref{eq:xirbtz}) deviates significantly from the simulation results (black), multi-sheath model (dashed red), and single-sheath model (dashed green). In fact, Eq.~(\ref{eq:xirbtz}) will always predict an ion channel with a half-length $L_h \equiv \xi(r_b = 0) - \xi(r_b = r_m) \approx 0.85 r_m$ whereas the bubble actually traces a nearly spherical shape $L_h \approx r_m$ when the blowout radius is large, i.e., $r_m\gtrsim 4$.

In Ref. \citenum{lu2006nonlinearphysplasma}, it was pointed out that the deviation between the analytic expression for $r_b(\xi)$ in Eq.~(\ref{eq:xirbtz}) and the actual wake trajectory from PIC simulations could be largely attributed to the additional $(dr_b/d\xi)^2$ term in Eq.~(\ref{eq:rbtz}) which caused the particle trajectories to bend toward the $\xi$-axis sooner than expected. Since the analytic theory underestimates the length of the ion channel, it naturally follows that as shown in Fig~\ref{fig:tzoufras_diff}(b) the electric field predicted by Eq.~(\ref{eq:efieldtzunloaded}) deviates from the empirical wakefield in a similar fashion. However, Eq.~(\ref{eq:efieldtzunloaded}) still captures the negative spike in the electric field near the axis since the slope of the trajectory $\frac{dr_b}{d\xi} = - \sqrt{\frac{r_m^4}{2r_b^4} - \frac{1}{2}}$ approaches $-\infty$ as $r_b \rightarrow 0$. 

Upon inspection of Eq.~(\ref{eq:drb}), this behavior arises because $P_{\perp}$ [Eq.~(\ref{eq:drb})] asymptotes to $-\infty$ since $\psi_{min} = 0$ for the wake potential $\psi_0 \approx r_b^2/4$. Thus, althougth its underlying approximations break down as $r_b \rightarrow 0$, the analytic model still predicts a spike because $P_{\perp} \rightarrow -\infty$ while the multi-sheath model predicts a spike because $(1+\psi) \rightarrow 0$. On the other hand the single-sheath model cannot predict a spike because $P_{\perp}$ remains finite and $\psi_{min} = 0$. This is perhaps the most important distinction between the analytic theory and single-sheath model in which the electric field is not a monotonically decreasing function of $\xi$ despite the fact that it also employs $\psi_{min} = 0$. As we have shown in Sec.~\ref{sec:beamloading}, this limitation of the single-sheath model at the rear of the wake is the primary reason why it cannot be used to design beam loads that produce constant wakefields, i.e., $\frac{dE_z}{d\xi} \simeq 0$. 

Despite the fact that Eqs.~(\ref{eq:xirbtz}) and (\ref{eq:efieldtzunloaded}) cannot accurately model the bubble radius and electric field as a function of $\xi$, the $(E_z,r_b)$ phase space predicted by Eq.~(\ref{eq:efieldtzunloaded}) agrees well with the simulation results and multi-sheath model for values of $r_b\gtrsim2$ as depicted in Fig.~\ref{fig:tzoufras_diff}(c). This is important because the analytic current profile described by Eq.~(\ref{eq:lambdaopt_1}) for loading a constant wakefield $E_z(\xi \geq \xi_t) = -E_t$ samples the phase space of $E_t$ and $r_t$ at the head of the load $\xi_t$. While Eqs.~(\ref{eq:xirbtz})-(\ref{eq:efieldtzunloaded}) do not accurately predict $r_t$ and $E_t$ as a function of $\xi_t$, Eq.~(\ref{eq:lambdaopt}) can be evaluated using the simulation data instead. Sampling the parameters this way will still produce self-consistent results in regions where the analytic theory is assumed to be valid $(r_b \gtrsim 3)$ since we are only shifting our initial position up the phase space curve $(E_z,r_b)$. 

For the profile described by Eq.~(\ref{eq:lambdaopt}), the loaded bubble trajectory obtained from Eq.~(\ref{eq:rbtz}) is parabolic $r_b^2  =r_t^2 -E_t(\xi-\xi_t)$. As we will show below, this expression underestimates the loaded wake length in the same manner as with Eq.~(\ref{eq:xirbtz}). Therefore, in many cases, the current profile predicted by analytic theory [Eq.~(\ref{eq:lambdaopt})] does not produce a perfectly constant wakefield over the entire bunch length. This can be seen in several examples provided in Ref.~\citenum{tzoufrasprab} where wakefields loaded using Eq.~(\ref{eq:lambdaopt}) still exhibit marginally nonzero slopes.

\begin{figure}[t]
\includegraphics[width=0.5\textwidth]{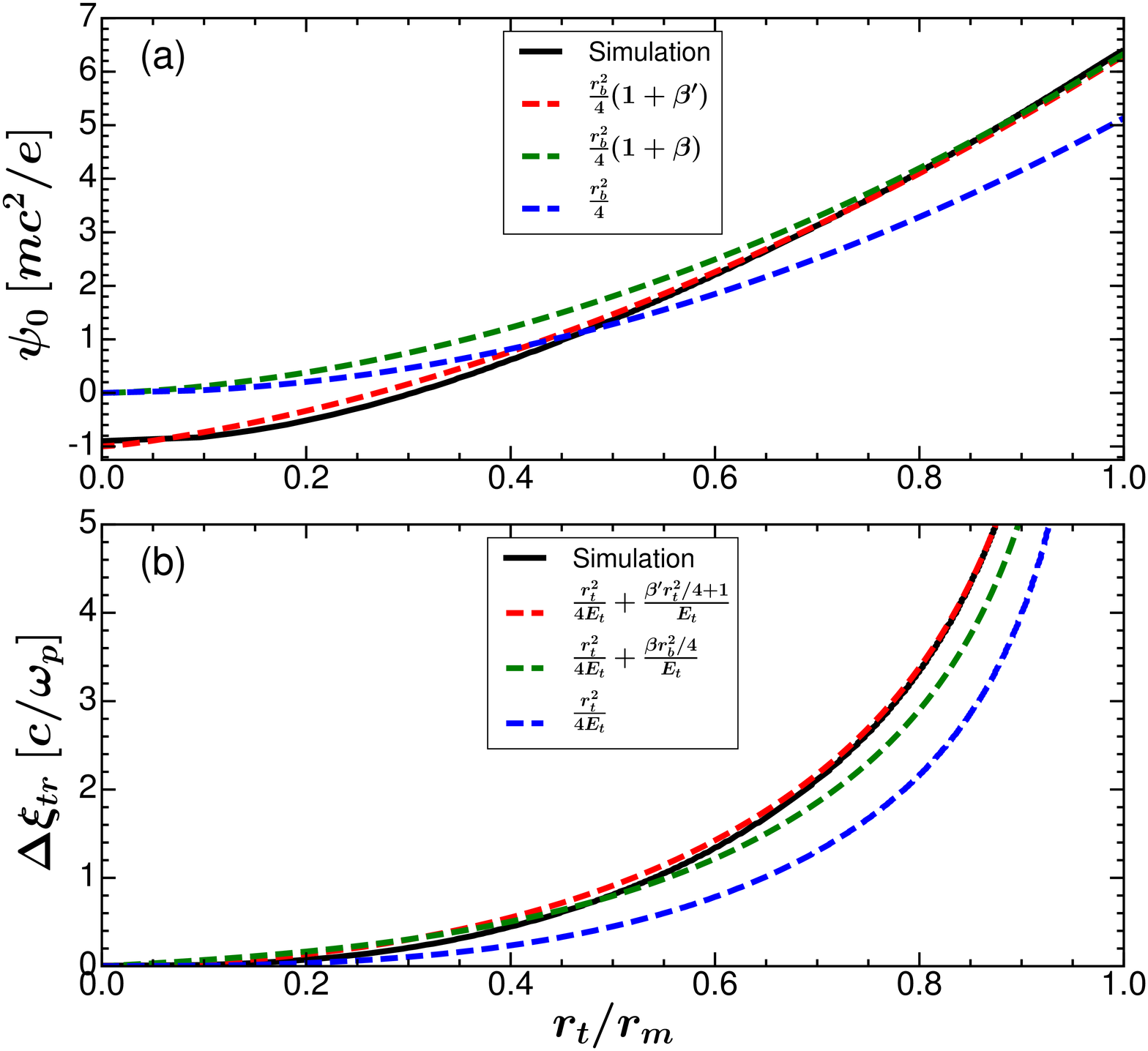}
\caption{\label{fig:xi_tr} (a) The on-axis wake potential $\psi_0$ and (b) maximum bunch length $\Delta \xi_{tr}$ [Eq.~(\ref{eq:deltaxi})] for loading a constant accelerating field $E_z(r_b \leq r_t) \simeq \text{const} \equiv -E_t$ as a function of $r_t/r_m$ for the plasma wake shown in Fig.~\ref{fig:tzoufras_diff}. The black curves use simulation data for $\psi(r_t)$ and $E_z(r_t)$ to evaluate Eq.~(\ref{eq:deltaxi}). Numerical calculations are shown for the multi-sheath model $\beta^{\prime}$ (dashed red) with $\psi_{min} = -1$ and the single-sheath model $\beta$ (dashed green) with $\psi_{min} = 0$ using Eqs.~(\ref{eq:psi_final})-(\ref{eq:n2_limit}) with $\Delta_{10} = 0.875$, $\Delta_s = 0.05r_b$, $\Delta_{20} = 3$, and $s =3$. The analytic theory (dashed blue) for $\psi_0 \approx \frac{r_b^2}{4}$ and $\Delta \xi_{tr} = \frac{r_t^2}{4E_t}$ is evaluated using simulation data for $r_t$ and $E_t$. }
\end{figure}

The disagreement between the analytic theory $(\beta, \beta^{\prime} \rightarrow 0,0)$ and multi-sheath model for $r_b(\xi)$ stems directly from the underlying assumption that the wake potential contributions from regions outside the bubble can be neglected, i.e., $\psi_0(\xi) \approx \Psi_{\text{I}} = \frac{r_b^2}{4}$. In Fig.~\ref{fig:xi_tr}(a), we compare the ion term $\frac{r_b^2}{4}$ (dashed blue) to $\psi_0$ obtained from the simulation results (black), the single-sheath model $\frac{r_b^2}{4}(1+\beta)$ (dashed green), and the multi-sheath model $\frac{r_b^2}{4}(1+\beta^{\prime})$ (dashed red) for the driver specified in Fig.~\ref{fig:tzoufras_diff}. 

From the results, it is clear that the analytic curve $(\beta^{\prime} \rightarrow 0)$ is only close to the simulation and multi-sheath model in a small region around $r_b \sim 0.45 r_m$. This is because the sheath components $\Psi_{\text{II}} + \Psi_{\text{III}} = \beta^{\prime}r_b^2/4$ cancel each other nearly exactly in this region as seen in Fig.~\ref{fig:example}(c). In contrast, the term $\beta r_b^2/4$ from the single-sheath model is positive-definite and only goes to zero when $r_b$ is zero. While the ion contribution $\Psi_{\text{I}} = r_b^2/4$ is the leading term when $r_b \approx r_m$, it underestimates the wake potential since the sheath term $\Psi_{\text{II}}(r_m)$ is on the order of unity while $\Psi_{\text{III}}(r_m)$ is approximately zero, and can be neglected (since $n_2 \approx 0$) as shown in Figs.~\ref{fig:example}(b)-(c). In this region, the wake potentials of the single-sheath and multi-sheath models are in agreement with the simulation results since $\beta^{\prime} \approx \beta $. 

Even though the blowout radius $r_m \simeq 4.53$ is large, the ion term $r_b^2/4$ still underestimates the simulated wake potential $\frac{\psi_0(r_m) -r_m^2/4}{\psi_0(r_m)}$ by approximately $(\sim 20\%)$ at the top fo the bubble. Therefore, from Eq.~(\ref{eq:drb}), the bubble trajectory predicted by the analytic theory should initially bend toward the axis with a more negative slope resulting in a shorter wake length. 

On the other hand, when $r_b \ll r_m$, the ion term $\Psi_{\text{I}}$ can be neglected while $\Psi_{\text{II}}$ and $\Psi_{\text{III}}$ from the multi-sheath model are both negative at the rear of the wake and, when combined, capture the limiting behavior of the wake potential $\psi_{min} \equiv \lim_{r_b \rightarrow 0} \beta^{\prime}r_b^2/4  \approx -1$ in Fig.~\ref{fig:example}(c). However, the analytic theory using $\psi_0 \approx r_b^2/4$ and single-sheath model using $\psi_0 = (1+\beta)r_b^2/4$ result in $\psi_{min} = 0$ at the axis. In addition, it is evident that the sheath term $\beta^{\prime} \rightarrow -\infty$ near the axis which violates the underlying assumption of the analytic theory that $\beta^{\prime}$ can be neglected. When including the source terms outside the ion channel described by Eq.~(\ref{eq:sprofile}) of the multi-sheath model, the wake potential $\psi_0=(1+\beta^{\prime})r_b^2/4$ obtained in Eq.~(\ref{eq:psi_final}) exhibits significantly improved agreement with the empirical simulation results across all values of $r_b$. As seen in Fig.~\ref{fig:xi_tr}(a), the curves for $\psi_0$ from the single and multi-sheath models deviate from each other at $r_t/r_m$ as large as $0.7$. This occurs because the curves must diverge such that $\psi_{min} = 0$ (single-sheath) rather than $\psi_{min} = -1$ (multi-sheath) for $r_t \rightarrow 0$.

The maximum theoretical length $\Delta \xi_{tr}$ over which the plasma wake can be loaded also depends on the profiles used for the on-axis wake potentials $\psi_0(r_b(\xi))$. In general, an expression for $\Delta \xi_{tr}$ can be obtained by integrating the electric field starting from the head of the bunch $r_b(\xi_t)=r_t$ to the rear of the bubble where the innermost particles cross the axis $r_b(\xi_f) = 0$ as follows
\begin{align}
\label{eq:deltaxi_integral}
\psi_0(r_t) - \psi_0(0) = -\int^{\xi_f }_{\xi_t} E_z d\xi
\end{align}
where $\xi_f = \xi_t + \Delta \xi_{tr}$. For a constant loaded wakefield $E_z(\xi_t \leq \xi \leq \xi_f) \equiv -E_t$, we obtain
\begin{align}
\label{eq:deltaxi}
\Delta \xi_{tr} = \frac{\psi_0(r_t) - \psi_{min}}{E_t}
\end{align}
where the minimum wake potential is defined by $\psi_{min} \equiv \psi_0(0)$. While the presence of the load will not modify the potential at the head $\psi_0(r_t)$ due to continuity with the unloaded region, it can alter the exact value of potential $\psi_{min}$ at the back of the bubble in some cases. For the purpose of this analysis, we will use $\psi_{min} \simeq -1$ to obtain an upper bound on $\Delta \xi_{tr}$ for the multi-sheath model.

We can now calculate the maximum bunch length $\Delta \xi_{tr}$ over which a constant wakefield can be loaded for the potential profiles specified in the analytic theory, single-sheath and multi-sheath models. For the wake potential $\psi_0(r_b) \approx \Psi_{\text{I}} = \frac{r_b^2}{4}$ from Eq.~(\ref{eq:psitz}), we recover the expression $\Delta \xi_{tr} = \frac{r_t^2}{4E_t}$. This expression was also derived in Sec.~\ref{sec:beamloading} by solving the loaded parabolic trajectory $\tilde{r}_b(\xi)^2 = r_t^2 - 4E_t(\xi- \xi_t)$ for $\tilde{r}_b = 0$ of the analytic theory. For the potential used in the single-sheath model $\psi_0(r_b) = \frac{r_b^2}{4}(1+\beta)$ where $\psi_{min} = 0$, the maximum bunch length is $\Delta \xi_{tr} =\frac{r_t^2}{4E_t} + \frac{\beta(r_t) r_t^2/4}{E_t}$. For the multi-sheath potential $\psi_0(r_b) = \frac{r_b^2}{4}(1+\beta^{\prime})$, where $\beta^{\prime}$ satisfies the condition $\psi_{min} = -1$, the maximum bunch length is $\Delta \xi_{tr} =\frac{r_t^2}{4E_t} + \frac{\beta^{\prime}(r_t) r_t^2/4 +1}{E_t}$. 

In Fig.~\ref{fig:xi_tr}(b), we plot $\frac{r_t^2}{4E_t}$ (dashed blue) calculated by extracting $(E_t,r_t)$ from simulation data along with $\Delta \xi_{tr}$ calculated from Eqs.~(\ref{eq:psi_final})-(\ref{eq:n2_limit}) for the single-sheath $\beta$ (dashed green) and multi-sheath model $\beta^{\prime}$ (dashed red). For reference, we also plot Eq.~(\ref{eq:deltaxi}) using values of $\psi_0$ and $E_t$ from simulation data (black). Since the potentials $\psi_0(r_t)$ in Fig.~\ref{fig:xi_tr}(a) are monotonically increasing, the maximum bunch length $\Delta \xi_{tr}$ that can be loaded increases with $r_t$ in each case. The limiting behavior $\Delta \xi_{tr} \rightarrow \infty$ is also observed in each case at the top of the bubble ($r_t = r_m$) where $\frac{dr_b}{d\xi} =0$ and $E_z(r_m)= -E_t = 0$. 

From the results displayed in Figs.~\ref{fig:xi_tr}(a)-(b), it can be readily seen that the multi-sheath model generates values of $\Delta \xi_{tr}$ that agree well with those calculated from simulation data while the model for $\psi_0$ used by Tzoufras et al.~\cite{tzoufrasprl, tzoufrasprab} underestimates the maximum bunch length for all values of $r_t$. The underlying reason is that the wake potential contributions $\Psi_{\text{II}}$ and $\Psi_{\text{III}}$ from source terms outside the ion channel are monotonically increasing with $r_b$ as shown in Fig.~\ref{fig:example}(c) and, therefore, add to the potential difference between any two points in the back half of the bubble. This potential difference manifests itself in the term $\frac{\beta^{\prime}(r_t) r_t^2/4 - \psi_{min}}{E_t}$, which is positive definite since $ \beta^{\prime}(r_t) r_t^2/4 = \Psi_{\text{II}} + \Psi_{\text{III}} \geq \psi_{min}$ for all $r_t$ as depicted in Fig.~\ref{fig:example}(c). Since the ion channel ends at the back of the bunch, i.e., $\xi_f= \xi_t + \Delta \xi_{tr}$, it also follows that the analytic expression for the bubble trajectory $r_b^2 = r_t^2 -E_t (\xi - \xi_t)$ also underestimates the length of the ion channel $L_h$ regardless of where the load is placed. While the single-sheath model predicts longer bunch lengths than the analytic theory due to the additional sheath term $\frac{\beta(r_t) r_t^2/4}{E_t}$ which is positive definite, it still falls short of the multi-sheath model since it does not account for the negative wake potential $\psi_{min} \approx -1$ near the axis.

\section{Conclusions}

We have proposed a multi-sheath phenomenological model for describing the source term profile $S \equiv -\frac{1}{en_p}(\rho - J_z/c)$ of plasma wakefields excited by relativistic electron drivers in the nonlinear blowout regime. Using the multi-sheath model, a new expression for the wake potential $\psi(r,\xi)$ is obtained and then used to solve for the trajectory of the innermost sheath electron $r_b$ by integrating the equation of motion from the nonlinear blowout theory \cite{lu2006nonlinearphysplasma}. In cases with and without trailing bunches, we have shown that the bubble radius $r_b$, wake potential $\psi_0$, and electric field $E_z$ predicted by the multi-sheath model demonstrate significantly improved agreement with simulations results at the rear of the wake when compared to the results from the sheath model by Lu et al.~\cite{lu2006nonlinearphysplasma}. In addition, the model demonstrates the capability to predict plasma wakefields in cases where electrons are injected at the rear of the bubble. We have shown how the multi-sheath model can be used to design beams that can load a constant wakefield and have discussed differences between the predictions for beam loading based on the multi-sheath model and single-sheath model in the ultrarelativistic limit used by Tzoufras et al.~\cite{tzoufrasprab}. Two examples are also provided in which the multi-sheath model is used to load longitudinally varying wakefields. Finally, we examined the shortcomings of the multi-sheath model in cases with laser drivers and briefly outlined how the model can be adapted in future work.

\section*{Acknowledgements}

This work was supported by US NSF grant No. 1806046, US DOE grant No. DE-SC0010064, and FNAL sub award  544405. The simulations were performed on the National Energy Research Scientific Computing Center (NERSC), a U.S. DOE Facility at Lawrence Berkeley National Laboratory, and Hoffman2 at UCLA.

\bibliographystyle{apsrev4-1}
\bibliography{refs_thamine}
\end{document}